\documentclass[aps,prb,twocolumn]{revtex4}
\usepackage{amsfonts}
\usepackage{amsmath}
\usepackage{graphicx}
\usepackage{dsfont}
\usepackage{diagbox}
\usepackage{array}
\usepackage{amssymb}
\usepackage{bbm}
\usepackage{float}
\usepackage{subfigure}
\usepackage{url}
\usepackage{hyperref}

\newcommand{\PreserveBackslash}[1]{\let\temp=\\#1\let\\=\temp}
\newcolumntype{C}[1]{>{\PreserveBackslash\centering}p{#1}}
\newcolumntype{R}[1]{>{\PreserveBackslash\raggedleft}p{#1}}
\newcolumntype{L}[1]{>{\PreserveBackslash\raggedright}p{#1}}

\begin{document}

\title{Tensor network state approach to quantum topological phase
transitions and their criticalities of $\mathbb{Z}_2$ topologically ordered
states}
\author{Wen-Tao Xu$^{1}$ and Guang-Ming Zhang$^{1,2}$}
\affiliation{$^{1}$State Key Laboratory of Low-Dimensional Quantum Physics and Department
of Physics, Tsinghua University, Beijing 100084, China. \\
$^{2}$Collaborative Innovation Center of Quantum Matter, Beijing 100084,
China.}
\date{\today}

\begin{abstract}
Due to the absence of local order parameters, it is a challenging task to
characterize the quantum topological phase transitions between topologically
ordered phases in two dimensions. In this paper, we construct a
topologically ordered tensor network wavefunction with one parameter $%
\lambda $, describing both the toric code state ($\lambda =1$) and double
semion state ($\lambda =-1$). Via calculating the correlation length defined
from the one-dimensional quantum transfer operator of the wave function
norm, we can map out the complete phase diagram in terms of the parameter $%
\lambda $ and three different quantum critical points (QCPs) at $\lambda =0$%
, $\pm 1.73$ are identified. The first one separates the toric code phase
and double semion phase, while later two describe the topological phase
transitions from the toric code phase or double semion phase to the symmetry
breaking phase, respectively. When mapping the quantum tensor network
wavefunction to the exactly solved statistical model, the norm of the wave
function is identified as the partition function of the classical
eight-vertex model, and both QCPs at $\lambda=\pm 1.73$ correspond to the
eight-vertex model at the critical point $\lambda =\sqrt{3}$, while the QCP
at $\lambda =0$ corresponds to the critical six-vertex model. Actually such
a quantum-classical mapping can not yield the complete low-energy
excitations at these three QCPs. We further demonstrate that the full
eigenvalue spectra of the transfer operators without/with the flux
insertions can give rise to the complete quantum criticalities, which are
described by the (2+0)-dimensional free boson conformal field theories
(CFTs) compactified on a circle with the radius $R=\sqrt{6}$ at $\lambda =\pm%
\sqrt{3}$ and $R=\sqrt{8/3}$ at $\lambda =0$. From the complete transfer
operator spectra, the finite-size spectra of the CFTs for the critical
eight-vertex model are obtained, and the topological sectors of anyonic
excitations are yielded as well. Furthermore, for the QCP at $\lambda =0$, no anyon
condensation occurs, but the emerged symmetries of the matrix product
operators significantly enrich the topological sectors of the CFT spectra.
Finally, we provide our understanding on the (2+0)-dimensional conformal
quantum criticalities and their possible connection with the generic
(2+1)-dimensional CFTs for quantum topological phase transitions.
\end{abstract}

\maketitle

\section{Introduction}

Topological phases of matter and topological order become two important
novel concepts in current condensed matter physics. In the past decade, the
theoretical framework for topological order in two dimensions has been
gradually established. For example, some exactly solvable models with
topological orders have been constructed \cite%
{kitaevToricCode2003,kitaevHoneycomb}, a physical picture for understanding
generic non-chiral topologically ordered states has been proposed \cite%
{levin_string-net_2005}, the fixed-point wave functions for arbitrary
non-chiral topological order are formulated in terms of tensor networks \cite%
{GuTensorNetwork}, and the profound underlying mathematics classifying the
topological ordered phases have also been unveiled \cite{Zoo}. Albeit so
many inspiring successes for understanding of topologically ordered phases
are achieved, the topological phase transitions among these topologically
ordered phases are still lack of a unified theoretical description.
Moreover, the anyon condensation mechanism provides us a physical picture
for some topological phase transitions \cite{Bais2009}, however, it can not
be generalized to characterize all kinds of topological phase transitions.
So it is urgent to develop a systematic theoretical framework for quantum
topological phase transitions.

The defining feature of topologically ordered phases is the ground state
degeneracy that depends on the topology of the manifold in which the systems
live. Other important properties, such as the topological entanglement
entropy and braiding statistics of anyons, can be extracted from the ground
state wave functions as well \cite{MES}. It is generally believed that
almost all information about the topologically ordered phase is encoded in
the ground states. But a generic ground state wave function of the
interacting many-body systems is an entangled complex which is hard to be
obtained. However, the majority of many-body physical ground states satisfy
the area law theorem, the corresponding wavefunctions can be efficiently
expressed in terms of tensor network states \cite{verstraete-cirac}. With
such tensor network states, many properties of the topological phases, such
as entanglement entropy, entanglement spectrum \cite{ESofPEPS}, modular
matrices\cite{ModularMatrices} and all kinds of expectation values, can be
efficiently calculated numerically. In one dimension, from the exact
ground-state wave functions in the matrix product state representation, we
have successfully decoded the quantum criticalities of the topological phase
transitions from the symmetry protected topological (SPT) phases to their
adjacent trivial phase with the same symmetry\cite%
{Rao2014,RaoYang2016,RaoZhu2016,WangZhuZhang}.

For the non-chiral topologically ordered tensor network states in two
dimensions, the central object is the matrix product operator (MPO), which
acts on the virtual (auxiliary) degrees of freedom of the tensor networks
\cite{AnyonMPO,MPOforSPT}. The insertion of a MPO in the tensor network
states on a torus can transform the topological ground state into its
degenerate ground state, just like the \textquotedblleft Wilson
loop\textquotedblright\ operator. Actually we can use these MPOs to
determine the anyonic excitations contained in the topologically ordered
tensor network states\cite{AnyonMPO}. In principle, by properly
parameterizing the local tensors with a finite bond dimension, the tensor
network state wave function can characterize various topological ordered
phases and even the quantum critical points. Such a tensor network state
approach is thus more useful than directly studying the topological phase
transitions from the parameterized model Hamiltonian, because solving a
quantum Hamiltonian with local interactions is usually more difficult
particular in two dimensions.

In this paper, we start from a two-dimensional tensor network wave function
for $\mathbb{Z}_{2}$ SPT phases with a tuning parameter $\lambda $ (Ref.\cite%
{HuangWei1}), where the protecting $\mathbb{Z}_{2}$ symmetry is
defined by the MPO of the tensor network. By introducing $\mathbb{Z}_{2}$
gauge degrees of freedom into the $\mathbb{Z}_{2}$ SPT tensor network state,
we can construct a parameterized tensor network wave function with $\mathbb{Z%
}_{2}$ topological orders, which incorporates the toric code phase, double
semion phase, the symmetry breaking phase, as well as the phase transition
points between these phases. Via calculating the correlation length from the
one-dimensional quantum transfer operator of the wave function norm, we can
map out the complete phase diagram and identify three different quantum
critical points (QCPs) at $\lambda =0$, $\pm 1.73$, respectively. The first
one separates the toric code phase ($0<\lambda <1.73$) and the double semion
phase ($-1.73<\lambda <0 $), while later two describe the topological phase
transitions from the toric code or double semion phase to the symmetry
breaking phase ($|\lambda |>1.73$), respectively. Moreover, this one
parameter family of tensor network wave function can be mapped into the
partition function of the exactly solved eight-vertex model, where the
phases with negative parameter $\lambda $ merge into those phases with
positive parameter $\lambda $. Then the double semion phase becomes
equivalent to the toric code phase, and the QCPs at $\lambda \simeq \pm 1.73$
correspond to the critical point $\lambda =\sqrt{3}$ of the eight-vertex
model. But the QCP at $\lambda =0$ corresponds to the critical six-vertex
model, a special critical point of the eight-vertex model.

In order to investigate the complete quantum criticalities associated to the
topological phase transitions, the full spectra of the transfer operators
without/with the flux insertion are carefully analyzed at these three QCPs
separately, and we find that the static correlators at all these three QCPs
are characterized by the two-dimensional free boson conformal field theory
(CFT) compactified on a circle with compactified radii: $R=\sqrt{6}$ for
QCPs at $\lambda =\pm \sqrt{3}$ and $R=\sqrt{8/3}$ for QCP at $\lambda =0$.
Moreover, the phase transitions at $\lambda =\pm \sqrt{3}$ can be related to
the anyon condensation in the tensor network formalism\cite%
{Shadowofanyons,AnyonsCondensation,Condensation_driven,AnyonCondensation2},
and the finite-size spectrum of the critical eight-vertex model corresponds
to the charge $1$ sector, while the sector of charge $-1$ exhibits the
condensation of topological anyon excitations. However, the QCP between two
topologically ordered phases at $\lambda =0$ is quite unusual, where the
transfer operators without/with flux insertion acquire additional MPO
symmetries. Thus there is a significant enrichment in the structure of CFT
spectra with fractionalized anyonic excitations. Finally, we provide our
understanding on such (2+0)-dimensional conformal quantum criticalities for
quantum topological phase transitions in two dimensions\cite{ardonne,CQCP}.

This paper is organized as follows. In Sec.II, we construct the one
parameter family of tensor network wave function with two distinct $\mathbb{Z%
}_{2}$ topological orders in two dimensions. In Sec.III, the one-dimensional
quantum transfer operator is introduced from the tensor network wave
function norm, and from the calculations of the correlation length the
complete phase diagram is mapped out and three different QCPs are
identified. In Sec.IV, the quantum-classical correspondence is carefully
discussed and the positions of those QCPs are confirmed from the exactly
solved statistical models. In Sec.V, to fully derive the spectra of the CFTs
of these three QCPs, we carefully analyze the complete spectra of the
transfer operators without/with the flux insertions. Sec.VI provides our
understanding of the (2+0)-dimensional conformal QCPs, and their
relationship with the generic (2+1)-dimensional CFTs for quantum topological
phase transitions. Some related discussions are included in the Appendices.

\section{Tensor-network wave functions}

\subsection{Wave functions for the $\mathbb{Z}_{2}$ SPT phases}

\begin{figure}[tbp]
\includegraphics[width=9cm, trim=0 70 0 0, clip]{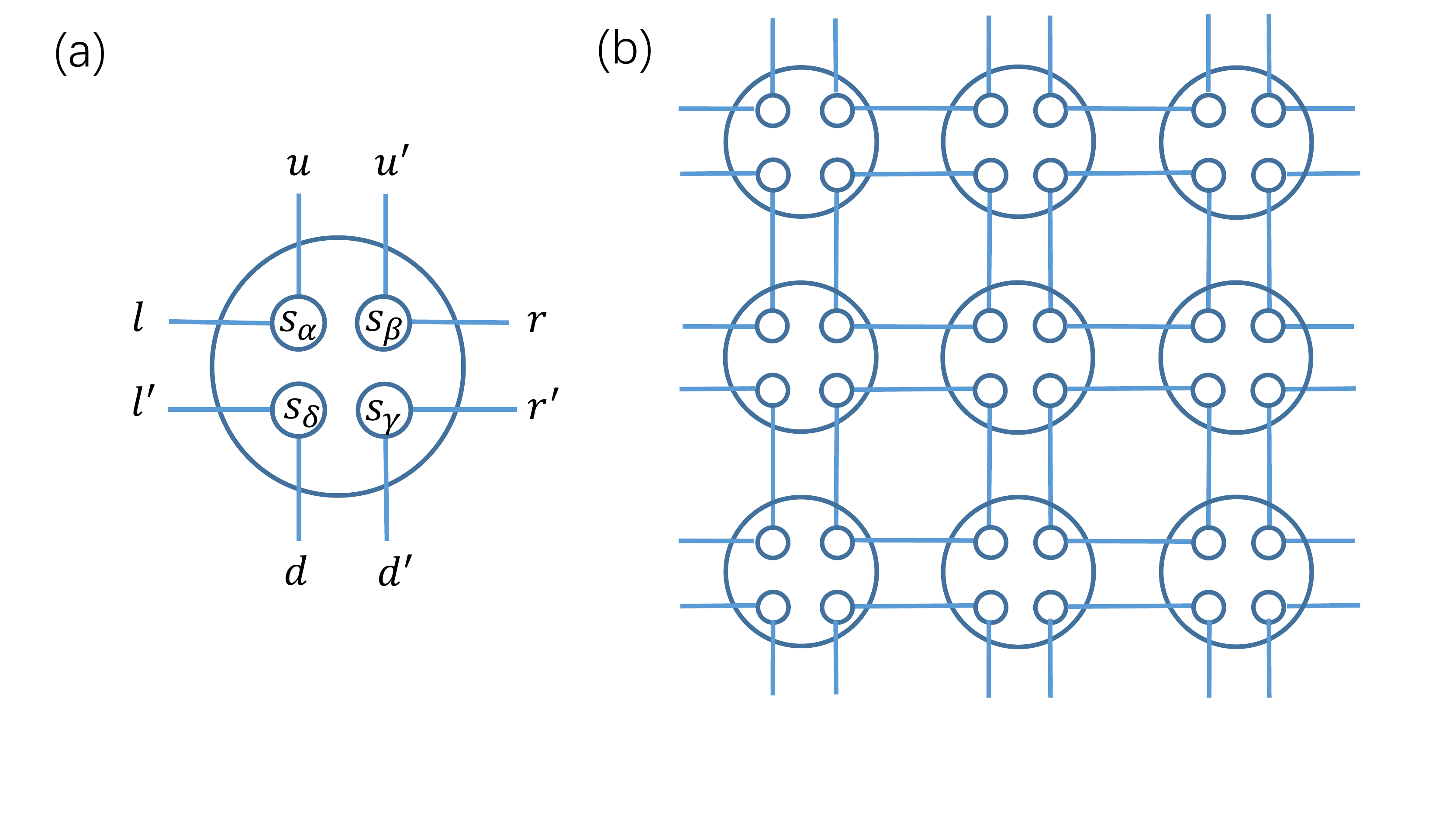}
\caption{(a) The structure of local tensor. (b) The SPT tensor network
state. }
\label{tensor}
\end{figure}

According to the classification theory for SPT phases in two dimensions\cite%
{SPT}, there exist two topologically distinct SPT phases protected by $%
\mathbb{Z}_{2}$ symmetry, one is trivial SPT phase and the other is
non-trivial SPT phase\cite%
{CZXmodel,ChiralSymmetry}. The fixed-point wave functions of these two
phases can be represented in terms of the double line tensor networks on a
two-dimensional square lattice\cite{CZXmodel,GuTensorNetwork}, as shown in
Fig. \ref{tensor}. Both physical and virtual degrees of freedom are $\mathbb{%
Z}_{2}$ spins. The physical indices of local tensor $\tilde{\mathcal{A}}$ are denoted by
the small circles, and a single lattice site represented by the big circle.
Four neighboring $\mathbb{Z}_{2}$ spins in the same inter-site plaquette
(the blue square in Fig. \ref{tensor}) are locked in the GHZ state $%
|0000\rangle +|1111\rangle $. Each local tensor $\tilde{\mathcal{A}}$ in
Fig. \ref{tensor}(a) has eight virtual indices ($u,u^{\prime },r,r^{\prime
},d,d^{\prime },l,l^{\prime }$) and four physical indices ($s_{\alpha
},s_{\beta },s_{\gamma },s_{\delta }$), and is given by
\begin{equation}
\tilde{\mathcal{A}}_{ll^{\prime }rr^{\prime }uu^{\prime }dd^{\prime
}}^{s_{\alpha }s_{\beta }s_{\gamma }s_{\delta }}=A^{s_{\alpha }s_{\beta
}s_{\gamma }s_{\delta }}\delta _{lu}^{s_{\alpha }}\delta _{u^{\prime
}r}^{s_{\beta }}\delta _{r^{\prime }d^{\prime }}^{s_{\gamma }}\delta
_{l^{\prime }d}^{s_{\delta }},  \label{SPTtensor}
\end{equation}%
where $\delta _{jk}^{i}=1$ for $i=j=k$ and $\delta _{jk}^{i}=0$ otherwise.
Because the entangled plaquettes are considered, the construction of double
line tensor network states intrinsically differs from the construction of
projected entangled pair states \cite{verstraete-cirac}, where only
entangled pair states are considered. Different tensor network wave
functions are determined by the choices of $A^{s_{\alpha }s_{\beta
}s_{\gamma }s_{\delta }}$. So the tensor network state is expressed by
contracting all adjacent local tensors:
\begin{equation}
|\psi (\lambda )\rangle =\sum_{\{s_{\alpha }\}}\text{tTr}(\tilde{\mathcal{A}}%
\otimes \cdots \otimes \tilde{\mathcal{A}})|s_{1}s_{2}s_{3}\cdots \rangle ,
\end{equation}%
where tTr denotes the tensor contraction over all virtual indices and $%
\{s_{\alpha }\}$ are all spin configurations. It has been proven that the
tensor network states can generally have a parent Hamiltonian as the
summation of local commuting projectors \cite{PEPS_as_ground_states}, so our
the tensor network state can be regarded as the ground state of a
two-dimensional quantum systems with local interactions.

\begin{figure}[tbp]
\includegraphics[width=9cm, trim=0 0 250 0, clip]{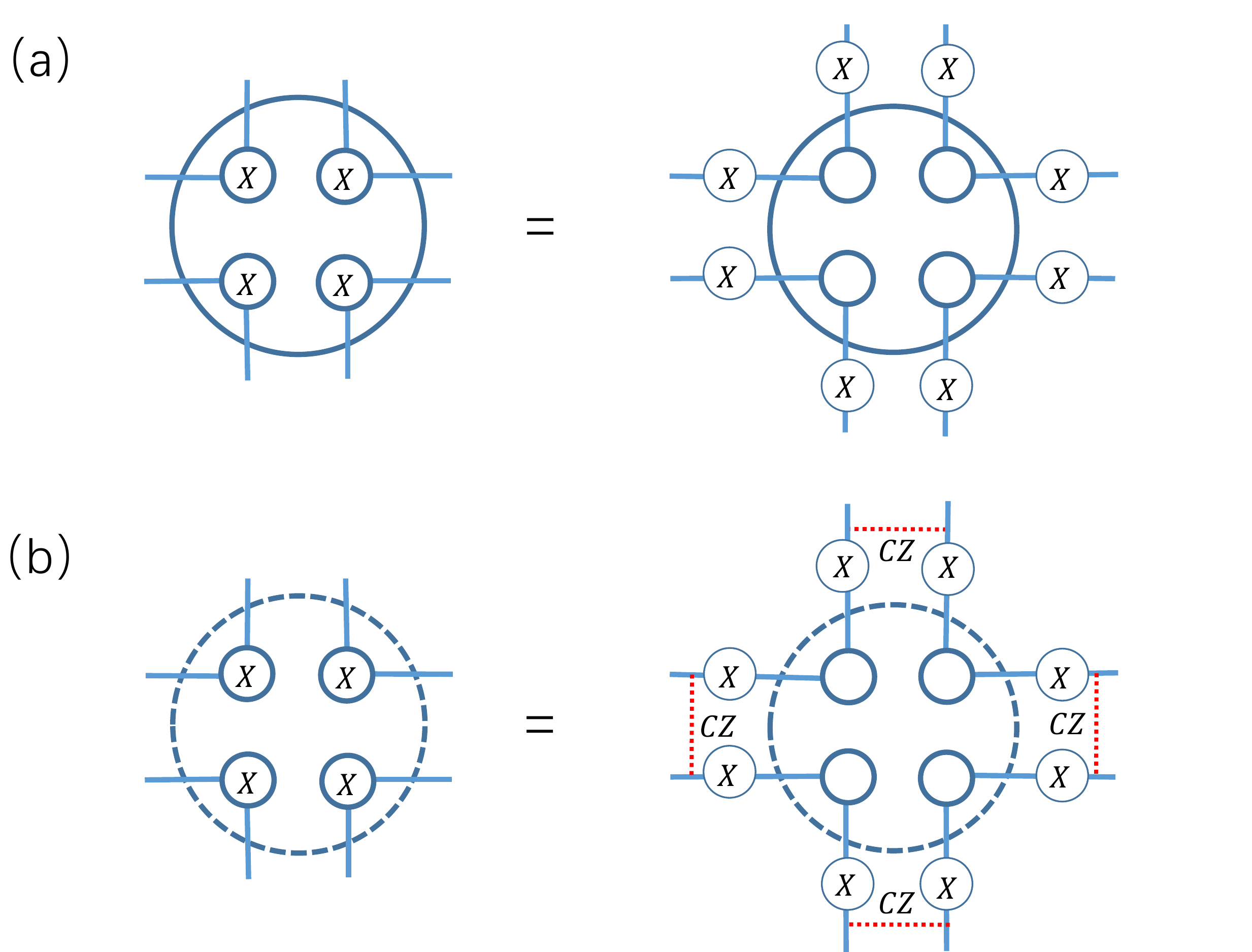}
\caption{Acting the on-site symmetry $X^{\otimes 4}$ on the physical degrees
of freedom is equivalent to acting MPO symmetry on the vritual degrees of
freedom. (a) The MPO for the local tensor $\tilde{\mathcal{A}} $ of the
trivial SPT state. (b) The MPO for local tensor of the nontrivial SPT state.}
\label{MPO}
\end{figure}

The main difference between the trivial and non-trivial SPT phases reflects
in the MPO representation of the on-site protecting symmetry on the virtual
space. For simplicity, the on-site protecting $\mathbb{Z}_{2}$ symmetry is
chosen as $\prod X^{\otimes 4}$, where $X^{\otimes 4}$ is the four Pauli
operators $\sigma ^{x}$ acting on the four spins on the same site. As shown
in Fig.\ref{MPO}, the protecting symmetry acting on the physical degrees of
freedom is equivalent to the MPO symmetry acting on the virtual degrees of
freedom\cite{MPOforSPT}. Note that the MPO representations of the on-site
symmetry on the virtual space are not unique, so we just follow the
conventional expression used in Ref.\cite{CZXmodel}. For the trivial SPT
phase, the MPO is trivial in the sense that the bond dimension of the MPO is
one, so it can be expressed as direct product of local operators: $U_{\text{x%
}}=\prod_{j=1}^{4}X_{j}^{\otimes 2}$, where $X_{j}^{\otimes 2}$ acts on the $%
j $-th quarter plaquette and the product runs over four adjacent quarter
plaquettes shown in Fig.\ref{MPO}(a). However, the bond dimension of MPO for
the non-trivial SPT phase is greater than one, and it is a non-on-site
operator given by\cite{CZXmodel}
\begin{equation}
U_{\text{czx}}=\prod_{j=1}^{4}X_{j}^{\otimes 2}\prod_{j=1}^{4}CZ_{j,j+1},
\end{equation}%
where $X_{j}^{\otimes 2}$ acts on the $j$-th quarter plaquette and $CZ_{j,j+1}=\text{%
diag}(1,1,1,-1)$ is the control-Z gate acting on the $j$-th and $(j+1)$-th
quarter plaquettes. When a quarter plaquette is grouped as a single degree
of freedom, one can find that the bond dimension of $U_{\text{czx}}$ is
indeed two. To satisfy the requirements of Fig.\ref{MPO}, the following
constraints on the local tensor have to be imposed:
\begin{eqnarray}
A^{0000} &=&A^{1111},A^{0001}=A^{1110},  \notag \\
A^{0010} &=&A^{1101},A^{0011}=A^{1100},  \notag \\
A^{0100} &=&A^{1011},A^{0101}=A^{1010},  \notag \\
A^{0110} &=&A^{1001},A^{0111}=A^{1000},\
\end{eqnarray}%
for the trivial SPT phase, and
\begin{eqnarray}
A^{0000} &=&A^{1111},A^{0001}=A^{1110},  \notag \\
A^{0010} &=&A^{1101},A^{0011}=-A^{1100},  \notag \\
A^{0100} &=&A^{1011},A^{0110}=-A^{1001},  \notag \\
A^{0101} &=&A^{1010},A^{0111}=A^{1000},
\end{eqnarray}%
for non-trivial SPT phase. Thus, it is the MPO symmetry that we can use to
distinguish the trivial or nontrivial SPT phases.

For convenience, the simplest local tensor $A$ for the trivial SPT tensor
network state is chosen as
\begin{equation}
A^{s_{\alpha }s_{\beta }s_{\gamma }s_{\delta }}=1,\ \text{for all}\
s_{\alpha }s_{\beta }s_{\gamma }s_{\delta },
\end{equation}%
while for non-trivial SPT tensor network state the local tensor is
\begin{equation}
A^{0011}=A^{0110}=-1; \ A^{s_{\alpha }s_{\beta }s_{\gamma }s_{\delta }}=1,\
\text{otherwise}.
\end{equation}%
Actually, these two tensor network states represent the fixed points of the
trivial and nontrivial SPT phases, respectively. More importantly, we can
construct a one parameter family of tensor network state with a
dimensionless tuning parameter $\lambda $ in the local tensor:
\begin{eqnarray}  \label{tensor_lambda}
A^{0011} &=&A^{0110}=\lambda ,  \notag \\
A^{1100}&=&A^{1001}=|\lambda |,  \notag \\
A^{s_{\alpha }s_{\beta }s_{\gamma }s_{\delta }} &=&1,\ \text{otherwise},
\label{aa}
\end{eqnarray}%
to incorporate both fixed-point wave functions $\lambda =\pm 1$
simultaneously \cite{HuangWei1}.

Moreover, for the general tensor network states, there exists a unitary
transformation
\begin{equation}
W=\prod \sum_{s_{\alpha }s_{\beta }s_{\gamma }s_{\delta }}w(s_{\alpha
}s_{\beta }s_{\gamma }s_{\delta })|s_{\alpha }s_{\beta }s_{\gamma }s_{\delta
}\rangle \langle s_{\alpha }s_{\beta }s_{\gamma }s_{\delta }|,
\end{equation}%
with $w(0011)=w\left( 0110\right) =-1$ and $w(s_{\alpha }s_{\beta }s_{\gamma
}s_{\delta })=1$ otherwise. Under this transformation, the tensor-network
state $|\psi (\lambda )\rangle $ is changed into $|\psi (-\lambda )\rangle $%
. Since such a transformation plays the role of a duality transformation, $%
\lambda =0$ becomes a self-dual point and the wave function $|\psi
(0)\rangle $ is invariant under this transformation. If the protecting
symmetries are preserved in the parameter space, tuning the parameter $%
\lambda $ from $1$ to $-1$, one can encounter a topological quantum phase
transition point between these two SPT phases. Exactly at the transition
point, the critical tensor network state has both MPO representations of $U_{%
\text{x}}$ and $U_{\text{czx}}$.

\begin{figure}[tbp]
\includegraphics[width=8.5cm]{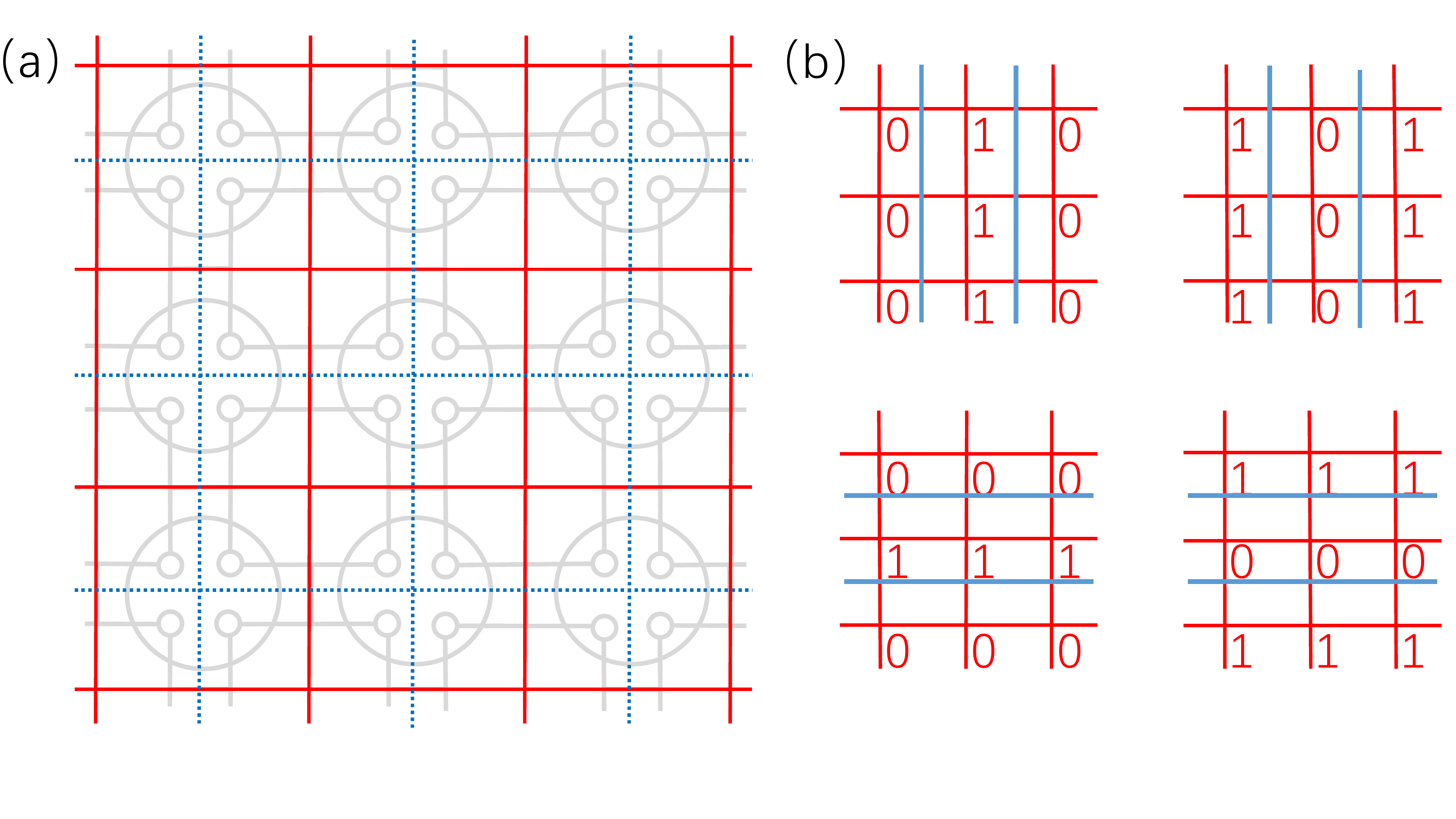}
\caption{(a) If the inter-site plaquette (gray square) is grouped as a
single spin at the vertices of the red lattice, the tensor network wave
functions can be further simplified, and the original lattice sites form the
blue square lattice. (b) On the red lattice, those four symmetry breaking
ground states for the tensor network state $|\protect\psi \rangle $ are
displayed, the numbers attached to the vertices of the red lattice are the
values of the $\mathbb{Z}_{2}$ spins, and the ground states have the
stripe-like ferromagnetic long-range order. The symmetry breaking ground
state configurations for $|\Psi \rangle $ are denoted by the blue lines,
which are the domain walls of spin configurations. }
\label{Blocking}
\end{figure}

Since those four spins in the same inter-site plaquette are locked with each
other, we can group the inter-site plaquette as a single spin located at the
vertices of the red lattice in Fig. \ref{Blocking}(a). When $|\lambda |\gg 1$%
, there only leave four dominant elements in the local tensor $A^{s_{\alpha
}s_{\beta }s_{\gamma }s_{\delta }}$, which form four different degenerate
ground states configurations for an arbitrary system size, corresponding to
the spontaneous symmetry breaking phase with the stripe ferromagnetic order,
as shown in Fig.\ref{Blocking} (b). Because the $\mathbb{Z}_{2} $ SPT phases
and symmetry breaking phase with the stripe ferromagnetic order have
different symmetries, there must exist two additional phase transition
points between these two SPT phases and symmetry breaking phase. The
detailed analysis of the phase diagram is given in the Appendix \ref%
{Phase_diagram_of_SPT}.

\subsection{Wave functions for $\mathbb{Z}_{2}$ topologically ordered phases}

\begin{figure}[tbp]
\includegraphics[width=9cm]{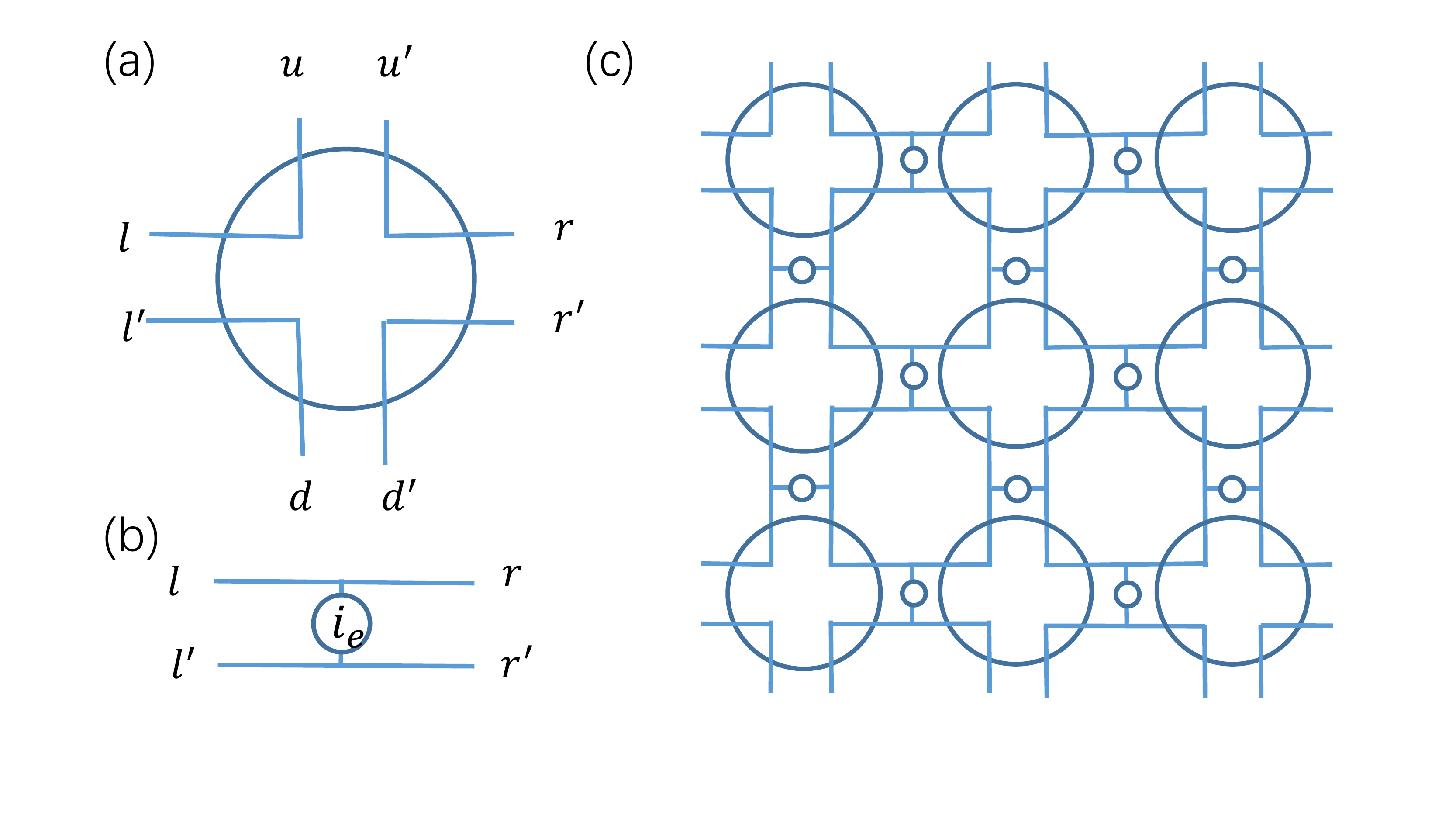}
\caption{(a) The local tensor $\mathcal{A}$ on the vertices of the lattice.
(b) The local tensor $\mathcal{D}$ on the edges of the lattice. (c) The
tensor network wave function for the $\mathbb{Z}_2$ topologically ordered
phases.}
\label{ToricCodeTensor}
\end{figure}

Since the $\mathbb{Z}_{2}$ topologically ordered states can be obtained by
gauging the $\mathbb{Z}_{2}$ SPT states \cite{LevinGu}, we construct the
tensor networks for $\mathbb{Z}_{2}$ topologically ordered states based on
the previous $\mathbb{Z}_{2}$ SPT tensor network states. At first we promote
the global $\mathbb{Z}_{2}$ symmetry to $\mathbb{Z}_{2}$ gauge symmetry by
introducing the $\mathbb{Z}_{2}$ gauge fields on two adjacent plaquettes,
see in Fig.\ref{ToricCodeTensor}(b), and the original $\mathbb{Z}_{2} $
spins are coupled to the $\mathbb{Z}_{2}$ gauge fields. Then integrating out
the physical degrees of freedom gives rise to the tensor network wave
functions for the $\mathbb{Z}_{2}$ topologically ordered phases. After the
gauging procedure, the trivial $\mathbb{Z}_{2}$ SPT state becomes the toric
code state, while the non-trivial SPT state is related to the double semion
state \cite{LevinGu}.

When the $\mathbb{Z}_{2} $ gauge fields are introduced with the local tensor
$\mathcal{D}$:
\begin{equation}
\mathcal{D}_{ll^{\prime }rr^{\prime }}^{i_{e}}=D_{ll^{\prime
}}^{i_{e}}\delta _{lr}\delta _{l^{\prime }r^{\prime }},
\end{equation}%
where $D_{10}^{1}=D_{01}^{1}=D_{00}^{0}=D_{11}^{0}=1$, otherwise $%
D_{ll^{\prime }}^{i_{e}}=0$. The physical degrees of freedom $i_{e}$ located
at the edges of the lattice are $\mathbb{Z}_{2} $ gauge field, and other
indices stand for the virtual degrees of freedom. The local tensor $\mathcal{%
D}$ actually detects the existence of the domain walls between the adjacent
inter-site plaquettes, as shown in Fig.\ref{ToricCodeTensor}(c). Including
the tensor $\mathcal{D}$ into the original SPT tensor network states leads
to the gauged SPT tensor network wave function:
\begin{equation}
|\Psi ^{\prime }\rangle =\sum_{\{i_{e}\}\{s_{\alpha }\}}\text{tTr}(\tilde{%
\mathcal{A}}\otimes \mathcal{D}\cdots \otimes \tilde{\mathcal{A}}\otimes
\mathcal{D})|i_{1}i_{2}\cdots \rangle |s_{1}s_{2}\cdots \rangle .
\end{equation}%
Then we can integrate out all $\mathbb{Z}_{2}$ spins by performing the
overlap $\prod \sum_{s_{\alpha }s_{\beta }s_{\gamma }s_{\delta }}\langle {%
s_{\alpha }s_{\beta }s_{\gamma }s_{\delta }}|\Psi ^{\prime }\rangle $, where
the bra state is the equal weight superposition of all spin configurations
and the product runs over all the lattice sites. And the tensor network wave
function for topologically ordered phases with only $\mathbb{Z}_{2}$ gauge
fields is thus obtained:\cite{ModularMatrices}
\begin{equation}
|\Psi (\lambda )\rangle =\sum_{\{i_{e}\}}\text{tTr}(\mathcal{A}\otimes
\mathcal{D}\cdots \otimes \mathcal{A}\otimes \mathcal{D})|i_{1}i_{2}i_{3}%
\cdots \rangle ,
\end{equation}%
where the local tensor $\mathcal{A}$ is given by
\begin{equation}
\mathcal{A}_{ll^{\prime }rr^{\prime }uu^{\prime }dd^{\prime }}=A^{lu^{\prime
}r^{\prime }d}\delta _{ul}\delta _{u^{\prime }r}\delta _{r^{\prime
}d^{\prime }}\delta _{l^{\prime }d},
\end{equation}%
displayed in Fig. \ref{ToricCodeTensor}, and the local tensor $A$ has been
defined by Eq. (\ref{aa}). Compared to the original local tensor $\tilde{%
\mathcal{A}}$, $\mathcal{A}$ does not include any physical indices.

Since almost all data of topological order are encoded in the modular
matrices which are determined by properties of MPOs, we can verify that the
tensor network wave function $|\Psi (\lambda =1)\rangle $ yields the
fixed-point tensor network state of the toric code model, while $|\Psi
(\lambda =-1)\rangle $ yields the fixed point tensor network state of the
double semion model. The detail analysis are given in the Appendix \ref%
{Modular_Matrices}. Unlike the unitary transformation $W$ for the SPT
states, the unitary transformation exchanging $|\Psi (\lambda)\rangle $ and $%
|\Psi (-\lambda )\rangle $ becomes non-local in the sense that it can not be
expressed as a product of local operators. However, this non-local unitary
transformation plays the similar role as the duality transformation from the
toric code phase to the double semion phase, so $\lambda =0$ is also the
self-dual point, describing the quantum topological phase transition between
two topologically ordered phases.

Similarly the wave function $|\Psi \left( \lambda \right) \rangle $ for $%
|\lambda |\gg 1$ also corresponds to the symmetry breaking phase with two
degenerate ground states: one consists of all vertical domain walls and the
other consists of all horizontal domain walls, as shown in Fig. \ref%
{Blocking}(b). Because two spin configurations correspond to one domain wall
configuration, the ground-state degeneracy of the symmetry breaking phase is
less than four. Two topological phases with topological orders are
significantly distinct from the symmetry breaking phase, we can expect the
presence of two additional phase transition points separating the two
topologically ordered phases from the symmetry breaking phase in the full
parameter space.

\begin{figure}[tbp]
\includegraphics[width=9cm]{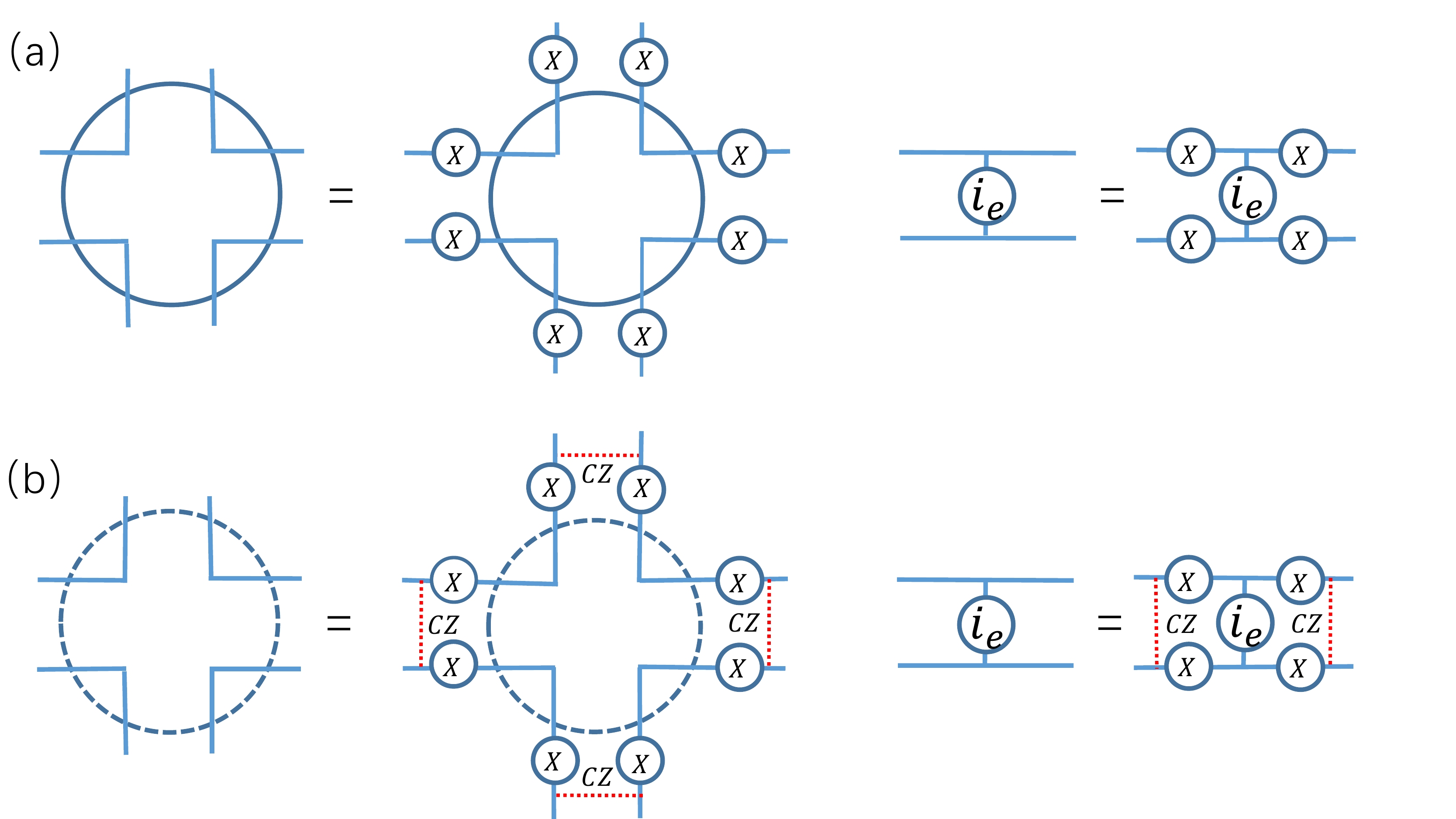}
\caption{The gauge symmetry of the local tensors of the topological states.
(a) For the toric code state, the local tensors $\mathcal{A}$ and $\mathcal{D%
}$ are invariant under the MPO action of $U_{\text{x}}$. (b) For the double
semion state, the local tensors are invariant under the MPO action the gauge
symmetry of $U_{\text{czx}}$. }
\label{MPOofTO}
\end{figure}

In addition, as shown in Fig. \ref{MPOofTO}, the MPOs for the respective SPT
phases now become the MPOs for the corresponding topologically ordered
phases, because the local tensors $\mathcal{A}$ and $\mathcal{D}$ are
invariant under the action of the the respective MPO. For the topologically
ordered phases, it has been known that the ground state degeneracy on a
torus is four, and these four degenerate wavefunctions correspond to the
present tensor network wave function without the MPO insertion, and with the
MPO insertion in either horizontal or vertical direction, or in both
directions, respectively. Moreover, the properties of these MPOs can also
give rise to the modular matrices for these two intrinsically $\mathbb{Z}_{2}
$ topological phases, which is crucial for identifying the characteristics
of topological order. The detail is also presented in the Appendix \ref%
{Modular_Matrices}.

\section{Quantum transfer operators and complete phase diagram}

\subsection{Quantum transfer operators}

Since the $\mathbb{Z}_{2}$ SPT states and the $\mathbb{Z}_{2}$ topologically
ordered states are associated by the gauging procedure, which preserves the
energy gap of the parent Hamiltonians of the tensor network states \cite%
{MPOforSPT}, we will expect that the structure of the phase diagram for the
topologically ordered states is similar to that for the SPT states \cite%
{HuangWei1}. The complete phase diagram of the general tensor-network with
topological order can be mapped out by calculating the correlation length as
a function of $\lambda $.

\begin{figure}[tbp]
\includegraphics[width=8cm,trim = 90 0 50
0,clip]{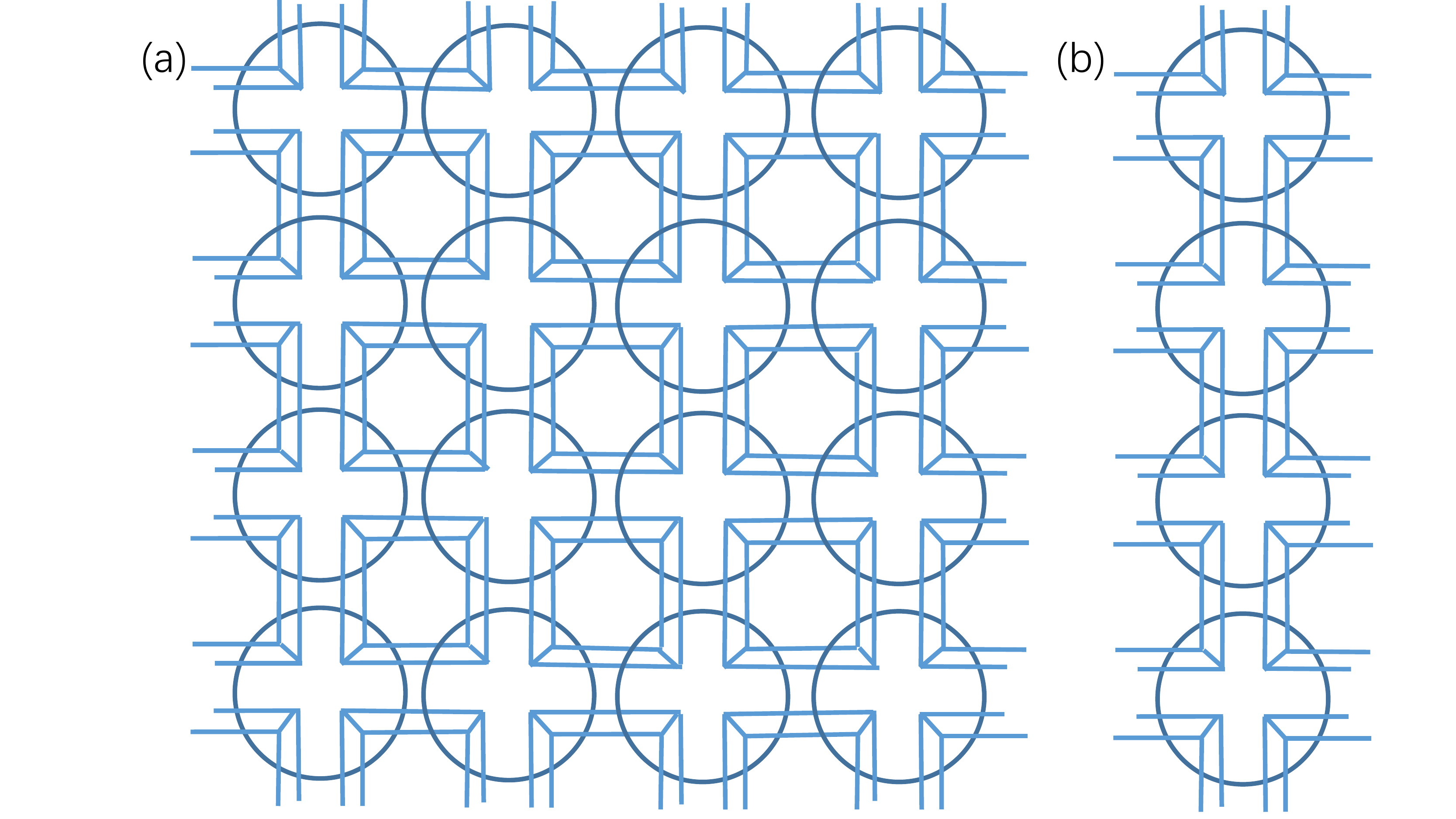}
\caption{(a) The double layer tensor network of wavefunction norm $%
\langle\Psi|\Psi\rangle$, where the left and right, up and down indices
should be contracted periodically. (b) The one-dimensional quantum transfer
operator, where the up and down indices should be connected periodically.}
\label{double_layer_tensor_network}
\end{figure}

In the tensor network representation, the correlation length can be
extracted from the dominant eigenvalues of the one-dimensional quantum
transfer operator $\mathbb{T}$, while the transfer operator is defined by
the tensor network wave function norm on a torus,
\begin{equation}
\langle \Psi |\Psi \rangle =\text{Tr}(\mathbb{T}^{N_{x}}),
\end{equation}
which is a double layer tensor network obtained by contracting the physical
indices of the tensor network (ket layer) and its complex conjugate (bra
layer), as displayed in Fig. \ref{double_layer_tensor_network}(a). The
one-dimensional quantum transfer operator is the repeating unit of the
double tensor network. Here $N_{x}$ is the number of lattice sites in the $x$
direction and the circumference of the transfer operator is given by $N_{y}$
in Fig. \ref{double_layer_tensor_network}(b). $N_{x}$ can be a very large
number while $N_{y}$ is limited by the numerical calculations.

In general, all kinds static correlation functions are closely related the
quantum transfer operator $\mathbb{T}$ (Ref.\cite{TransferMatrices}). Since
its eigenvalue spectrum contains essential information about the bulk
properties of systems\cite{TransferMatrices,Shadowofanyons}, such a 
one-dimensional quantum transfer operator can be viewed as a matrix with the 
left indices as the row and right indices as the column so that the numerical 
calculations can be performed. The transfer operator is
usually not necessarily hermitian, but we can take the modulus of the
eigenvalues. The finite degeneracy of the dominant eigenvalues indicates the
presence of either topological order or long-range order with a spontaneous
symmetry breaking. Therefore, the one-dimensional quantum transfer operator
plays the essential role in the present tensor network state approach. It is
the central object which we study in this paper.

The quantum transfer operator generally includes the bra and ket layer
structures. For the topological phases, the spin degrees of freedom in the
bra layer and the ket layer are connected by the same domain wall degrees of
freedom, so the spin configurations in both bra and ket layers are either
the same or opposite. Then the transfer operator can be decomposed into
\begin{equation}
\mathbb{T}=\mathbb{T}_{0}\oplus \mathbb{T}_{1},
\end{equation}%
where $\mathbb{T}_{0}$ is one subblock with the same spin configurations in
the bra and ket layers and $\mathbb{T}_{1}$ is the second subblock with the
opposite spin configurations in the bra and ket layers. Actually the
transfer operator also inherits the MPO symmetry of the local tensors, which
includes two $\mathbb{Z}_{2}$ symmetries acting on the bra and ket layers
respectively:
\begin{equation}
\mathbb{Z}_{2}\otimes \mathbb{Z}_{2}=\{\mathbbm{1}\otimes \mathbbm{1},%
\mathbbm{1}\otimes U_{\phi },U_{\phi }\otimes \mathbbm{1},U_{\phi }\otimes
U_{\phi }\},
\end{equation}%
where $\phi $ is `x' for $\lambda >0$ and `czx' for $\lambda <0$, and
\begin{equation}
U_{\text{x}}=\prod_{j}X_{j}^{\otimes 2},\text{ }U_{\text{czx}%
}=\prod_{j}X_{j}^{\otimes 2}\prod_{j}CZ_{j,j+1},
\end{equation}%
with $X_{j}^{\otimes 2}$ acts on the $j$-th half plaquette and $CZ_{j,j+1}$
acts on the $j$-th and $(j+1)$-th half plaquettes.

Since two subblocks $\mathbb{T}_{0}$ and $\mathbb{T}_{1}$ can be changed
into each other by the transformations $\mathbbm{1}\otimes U_{\phi }$ or $%
U_{\phi }\otimes \mathbbm{1}$, the eigenvalue spectra of $\mathbb{T}_{0}$
and $\mathbb{T}_{1}$ are the same. However, the eigenstates of $\mathbb{T}$
break the symmetry $\mathbbm{1}\otimes U_{\phi }$ and $U_{\phi }\otimes %
\mathbbm{1}$ but preserve the symmetry $U_{\phi }\otimes U_{\phi }$, so the $%
\mathbb{Z}_{2}\otimes \mathbb{Z}_{2}$ symmetry is broken down to $\mathbb{Z}%
_{2}$, leading to the exact two-fold degeneracy of the topological phases.
It has been established that different symmetry breaking patterns in the
virtual degrees of freedom of the transfer operator correspond to the phases
with distinct topological orders, so this kind of symmetry breaking pattern
just corresponds to the intrinsically $\mathbb{Z}_{2}$ topological phases
\cite%
{AnyonsCondensation,Shadowofanyons,AnyonCondensation2,Condensation_driven}.

As the unitary transformation exchanging the wave functions $|\Psi (\lambda
)\rangle $ and $|\Psi (-\lambda )\rangle $ acts on the physical degrees of
freedom, the transfer operators of $\mathbb{T}(\lambda )$ and $\mathbb{T}%
(-\lambda )$ are also related to each other by the unitary transformation
acting on the virtual degrees of freedom. Specially, the unitary
transformation acting on the subspace in which the degrees of freedom in the
bra and ket layers are opposite is given by $U_{\text{cz}}\otimes \mathbbm{1}
$ or $\mathbbm{1}\otimes U_{\text{cz}}$, where $U_{\text{cz}%
}=\prod_{j}CZ_{j,j+1}$. Then the eigenvalue spectra of $\mathbb{T}(\lambda )$
and $\mathbb{T}(-\lambda )$ are the same and symmetric about the self-dual
point $\lambda =0$, implying that the resulting transfer operator spectra
for both toric code phase and double semion phase are the same. As for the
differences between these two topological phases, we will discuss them in
the following sections, while the self-dual point $\lambda =0$ is very
intricate, where there is an emergent MPO symmetry $U_{\text{cz}}$.

\subsection{Transfer operators with flux insertions}

\begin{figure}[tbp]
\includegraphics[width=9cm, trim = 10 0 50
0,clip]{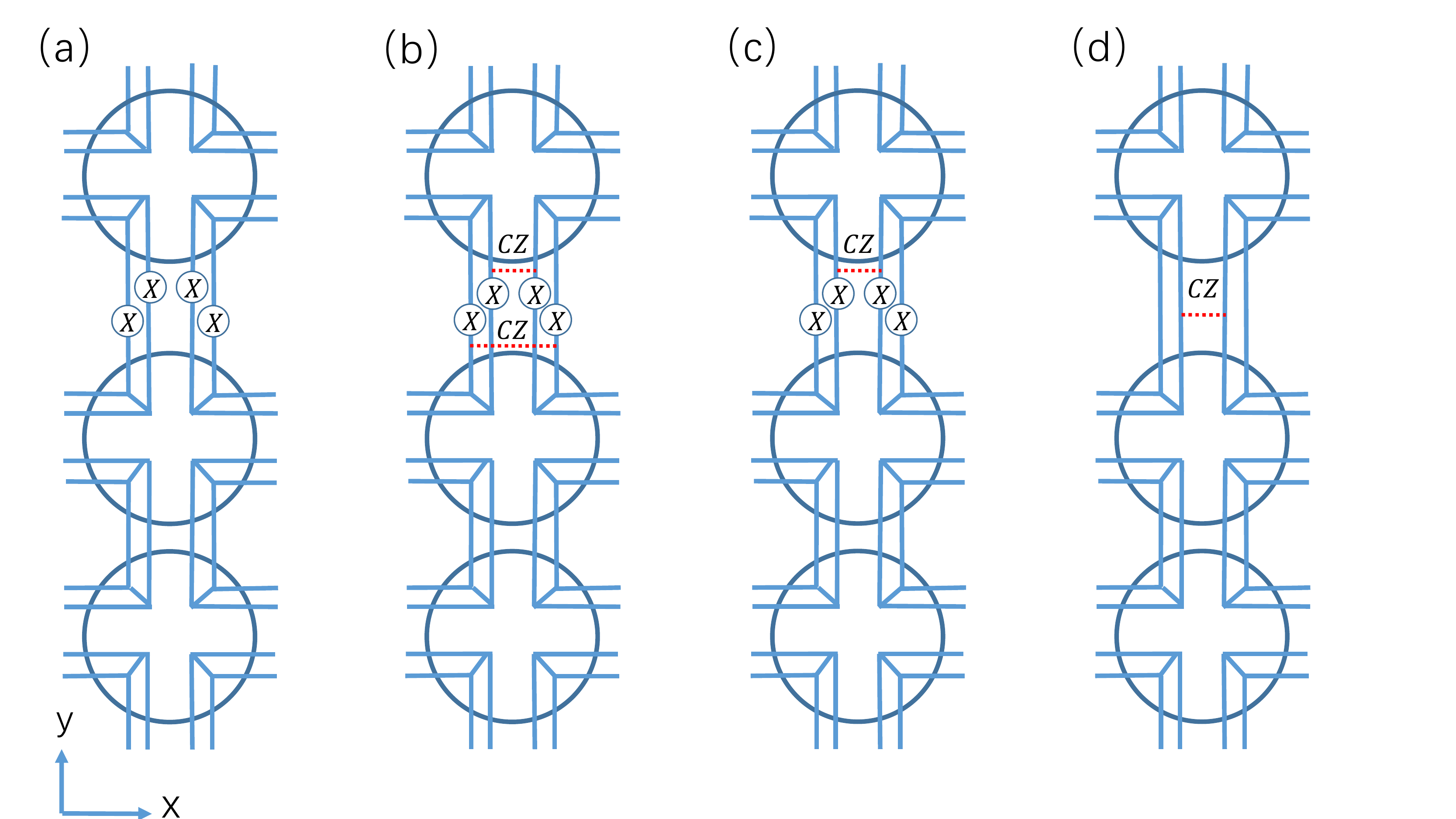}
\caption{The flux inserted transfer operators $\mathbb{T}_{\text{x}}^{\text{x%
}}$, $\mathbb{T}_{\text{czx}}^{\text{czx}}$, $\mathbb{T}_{\text{czx}}^{\text{%
x}}$,and $\mathbb{T}_{\text{cz}}$ are shown in (a), (b), (c) and (d),
respectively.}
\label{TO-flux}
\end{figure}

For the topologically ordered phases, the ground states are degenerate on a
manifold with non-trivial topology. In the tensor network formalism, these
degenerate ground states can be related by the MPO insertion in different
directions. Thus, the complete transfer operators should include the
transfer operator with different MPO insertions. For the topological phases,
the MPO operator $U_{\phi }$ can be inserted in the bra or ket layer of the
double layer tensor network, and the resulted transfer operator is denoted
as $\mathbb{T}^{\phi }$ or $\mathbb{T}_{\phi }$. However, such transfer
operators are not meaningful because they represent vanished overlaps
between different ground states. So we must insert the MPO operator $U_{\phi
}$ in both layers of the tensor networks, the resulting transfer operator is
denoted as $\mathbb{T}_{\phi }^{\phi }$ and shown in Figs.\ref{TO-flux}(a)
and (b). Inserting such a MPO flux into the transfer operator is equivalent
to the insertion of the MPO flux into $\mathbb{T}_{0}$ and $\mathbb{T}_{1}$
separately. So the operators $\mathbb{T}_{\phi }^{\phi }$ can be further
expressed as
\begin{equation}
\mathbb{T}_{\phi }^{\phi }=\mathbb{T}_{0,\phi }^{\phi }\oplus \mathbb{T}%
_{1,\phi }^{\phi }.
\end{equation}%
Similarly the transfer operator $\mathbb{T}_{\phi }^{\phi }$ also exhibits
the $\mathbb{Z}_{2}\times \mathbb{Z}_{2}$ symmetry. But the eigenstates of $%
\mathbb{T}_{\phi }^{\phi }$ break the $\mathbb{Z}_{2}\times \mathbb{Z}_{2}$
symmetry down to $\mathbb{Z}_{2}$ in the virtual degrees of freedom,
signifying the $\mathbb{Z}_{2}$ topological order with a two-fold degeneracy
in the eigenvalue spectrum. Meanwhile, the transfer operator $\mathbb{T}%
_{\phi }^{\phi }(\lambda )$ is also related to $\mathbb{T}_{\phi }^{\phi
}(-\lambda )$ by the unitary transformation, and their eigenvalue spectra
are the same and symmetric about $\lambda =0$. So we just need to calculate
the eigenvalue spectrum of $\mathbb{T}_{\text{x}}^{\text{x}}(\lambda )$ with
$\lambda >0$ in the following discussion.

By the way, we would like to point out that the transfer operators for the
intrinsically $\mathbb{Z}_{2}$ topological phases have a close relationship
with those transfer operators for the SPT phases. $\mathbb{T}_{0}$ is the
transfer operator of the tensor networks for the SPT phases, while $\mathbb{T%
}_{0,\phi }^{\phi }$ corresponds to the transfer operator of the SPT tensor
network with the extrinsic symmetric defect, which has been used to detect
the non-trivial SPT property \cite%
{SymmetryDefect,ABEffect,SCofSPT,Wang-Santos-Wen,MPOforSPT}.

\subsection{Transfer operators at $\protect\lambda =0$}

Different from the tensor network states for the finite values of $\lambda $%
, the tensor network state at $\lambda =0$ has both MPO symmetries $U_{\text{%
czx}}$ and $U_{\text{x}}$, because the topological properties for both sides
of the topologically ordered phases are inherited. Notice that MPOs of $U_{%
\text{x}}$ and $U_{\text{czx}}$ do not commute, so the gauge symmetry for
the tensor network state is not the simple $\mathbb{Z}_{2}\times \mathbb{Z}%
_{2}$ symmetry. Since there are two different MPO symmetries, we can insert
different MPOs in the bra and ket layers of the tensor networks, e.g., $%
\mathbb{T}_{\text{czx}}^{\text{x}}$ shown in Fig.\ref{TO-flux}(c). Actually,
the MPOs of $U_{\text{x}}$ and $U_{\text{czx}}$ in the same layer can be
fused into a new MPO of $U_{\text{cz}}$ when there is no MPO insertion in
the other direction. Then such a new MPO as the unitary transformation $%
\mathbbm{1}\otimes U_{\text{cz}}$ becomes an emergent symmetry of the
transfer operator. Inserting this new MPO in one of the bra and ket layers
of the double layer tensor networks gives rise to the twisted transfer
operator $\mathbb{T}_{\text{cz}}$, see Fig. \ref{TO-flux}(d). Although there
are other possible ways of inserting MPOs, they turn out to be either a
vanished transfer operator or the transfer operator whose spectrum is
equivalent to $\mathbb{T}$, $\mathbb{T}_{\text{x}}^{\text{x}}$, $\mathbb{T}_{%
\text{czx}}^{\text{x}}$ and $\mathbb{T}_{\text{cz}}$, respectively.

\subsection{Complete phase diagram}

The correlation length of the general tensor network state wave function $%
|\Psi (\lambda )\rangle $ can be calculated from the diagonalizing the
complete transfer operator $\mathbb{T}\oplus \mathbb{T}_{\phi }^{\phi }$
with the largest circumference of the transfer operators $N_{y}=20$ in the
numerical calculations. Each eigenvalue of $\mathbb{T}$ and $\mathbb{T}%
_{\phi }^{\phi }$ should have a two-fold degeneracy at least. In Fig.\ref{TO
correlation length}(a), we display the numerical results of the quantities $%
-1/\text{ln}|\frac{d_{i}}{d_{1}}|$, where $d_{i}$ is the $i$-th largest
eigenvalue of the transfer operator $\mathbb{T}\oplus \mathbb{T}_{\phi
}^{\phi }$. In the topological phases, the correlation length is determined
by $\xi =-1/\text{ln}|\frac{d_{5}}{d_{1}}|$, because the largest eigenvalue
has four-fold degeneracy in the topological phases. In the symmetry breaking
phase, however, $\xi =-1/\text{ln}|\frac{d_{9}}{d_{1}}|$, due to the
presence of an eight-fold degeneracy. For $|\lambda |\gg 1$, the present
tensor network description contains a redundancy in the symmetry breaking
phase: one domain wall configuration corresponds to two spin configurations.

According to the divergence of the correlation lengths shown in Fig. \ref{TO
correlation length}(a), we can thus identify three different QCPs at $%
\lambda =0$ and $\lambda =\pm 1.73$, dividing the phase diagram into three
different gapped phases: the toric code phase ($0<\lambda <1.73$), the
double semion phase ($-1.73<\lambda <0$), and the symmetry breaking phase ($%
|\lambda |>1.73$). The fixed point tensor network states of the toric code
and double semion models locate at $\lambda =\pm 1$ and have zero
correlation length.

\begin{figure}[tbp]
\includegraphics[width=8.5cm]{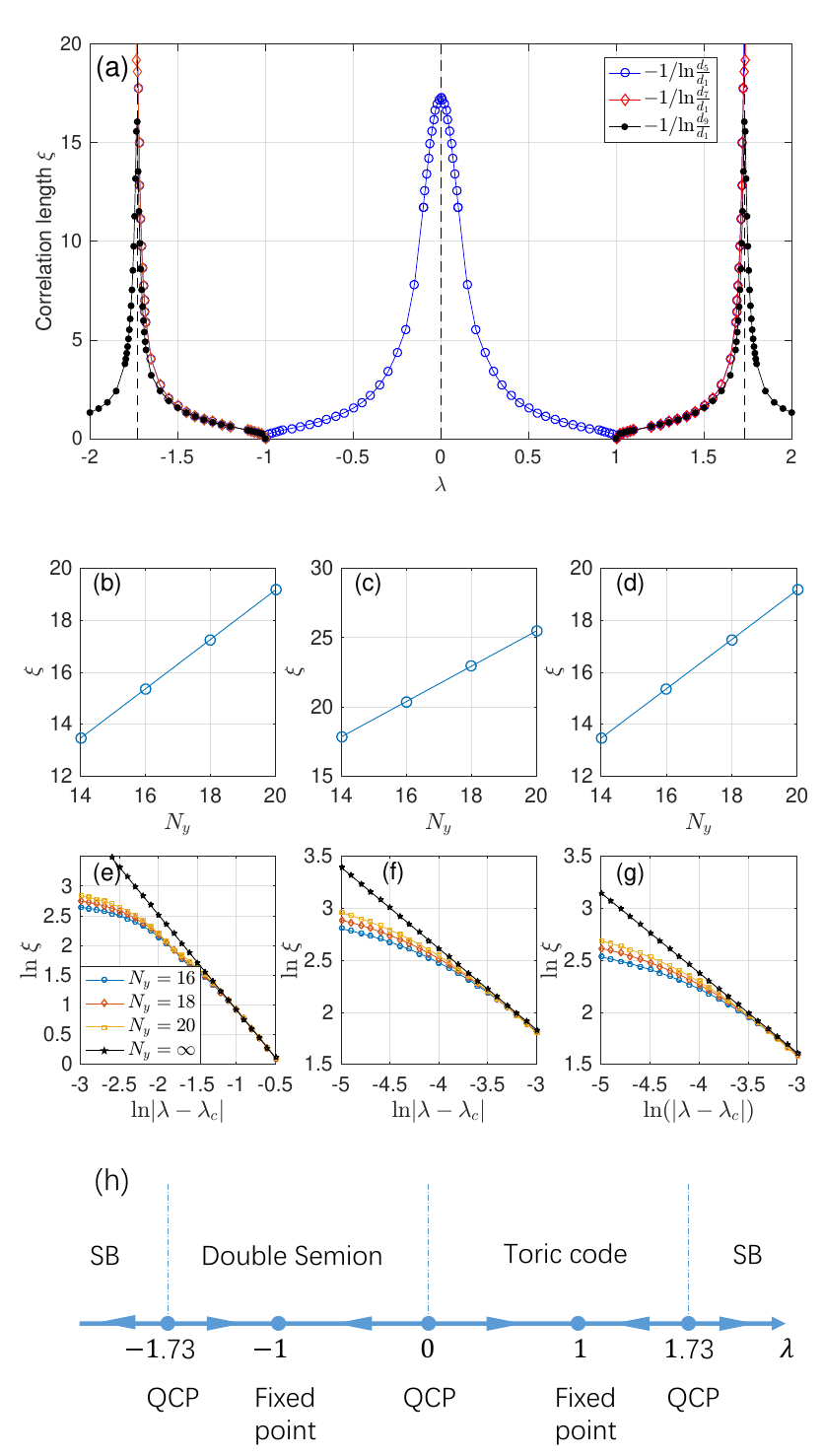}
\caption{(a) The quantities $-1/\text{ln}|\frac{d_{i}}{d_{1}}|$ as functions
of $\protect\lambda $, where $d_{i}$ is the $i$-th dominant eigenvalue of
the transfer operator $\mathbb{T}\oplus \mathbb{T}_{\text{x}}^{\text{x}}$.
Here the circumference of the transfer operator is chosen as $18$ sites.
(b), (c), (d) The finite correlation length satisfies $\protect\xi \propto
N_{y}$ at $\protect\lambda =0$, $\protect\lambda \approx 1.73^{-}$, and $%
\protect\lambda \approx 1.73^{+}$, respectively. (e),(f),(g) The critical
exponent $\protect\nu$ of the correlation length is fitted at $\protect%
\lambda \rightarrow 0$, $|\protect\lambda |\rightarrow 1.73^{-}$ and $|%
\protect\lambda |\rightarrow 1.73^{+}$. (h) The phase diagram is plotted and
several special points are marked, where the arrows denote the decreasing
direction of the correlation length as a function of the parameter $\protect%
\lambda$.}
\label{TO correlation length}
\end{figure}

In Fig. \ref{TO correlation length}(b), the correlation length in the
different lattice sizes satisfies $\xi \propto N_{y}$ at the phase
transition point $\lambda =0$, indicating that the correlation length tends
to infinite in the thermodynamic limit. So the topological phase transition
is continuous and the transition point is a QCP. In Fig. \ref{TO correlation
length}(e), the critical exponent of the correlation length is determined as
$\nu =\nu ^{\prime }\simeq 1.60$ for both $\lambda \rightarrow 0^{-}$ and $%
\lambda \rightarrow 0^{+}$. At $|\lambda _{c}|\approx 1.73$, however, the
correlation length is defined by $\xi =-1/\text{ln}|\frac{d_{5}}{d_{1}}|$
for $|\lambda |<1.73$, and $-1/\text{ln}(|\frac{d_{9}}{d_{1}}|)$ for $%
|\lambda |>1.73$, which are found to be proportional to the circumference of
the transfer operator $N_{y} $ as shown in Figs.\ref{TO correlation length}
(c) and (d), respectively. Thus these two phase transition points at $%
\lambda=\pm1.73$ are also QCPs. In Figs. \ref{TO correlation length} (f) and
(g), the critical exponents of the correlation length on both sides of the
transition point are fitted as $\nu \simeq 0.78$ for $|\lambda |\rightarrow
1.73^{-}$ and $\nu ^{\prime }\simeq 0.77$ for $|\lambda |\rightarrow 1.73^{+}
$, which are almost the same value. Since the critical exponent at $\lambda=0
$ is almost twice larger than that at the critical point $|\lambda|=1.73$,
the former correlation length looks less divergent than the latter.
Therefore, the full phase diagram is displayed in Fig. \ref{TO correlation
length}(h), where the arrows indicate the decreasing direction of the
correlation length, consistent with the renormalization group flow.

\begin{figure}[tbp]
\includegraphics[width=9cm, trim=170 150 170
150,clip]{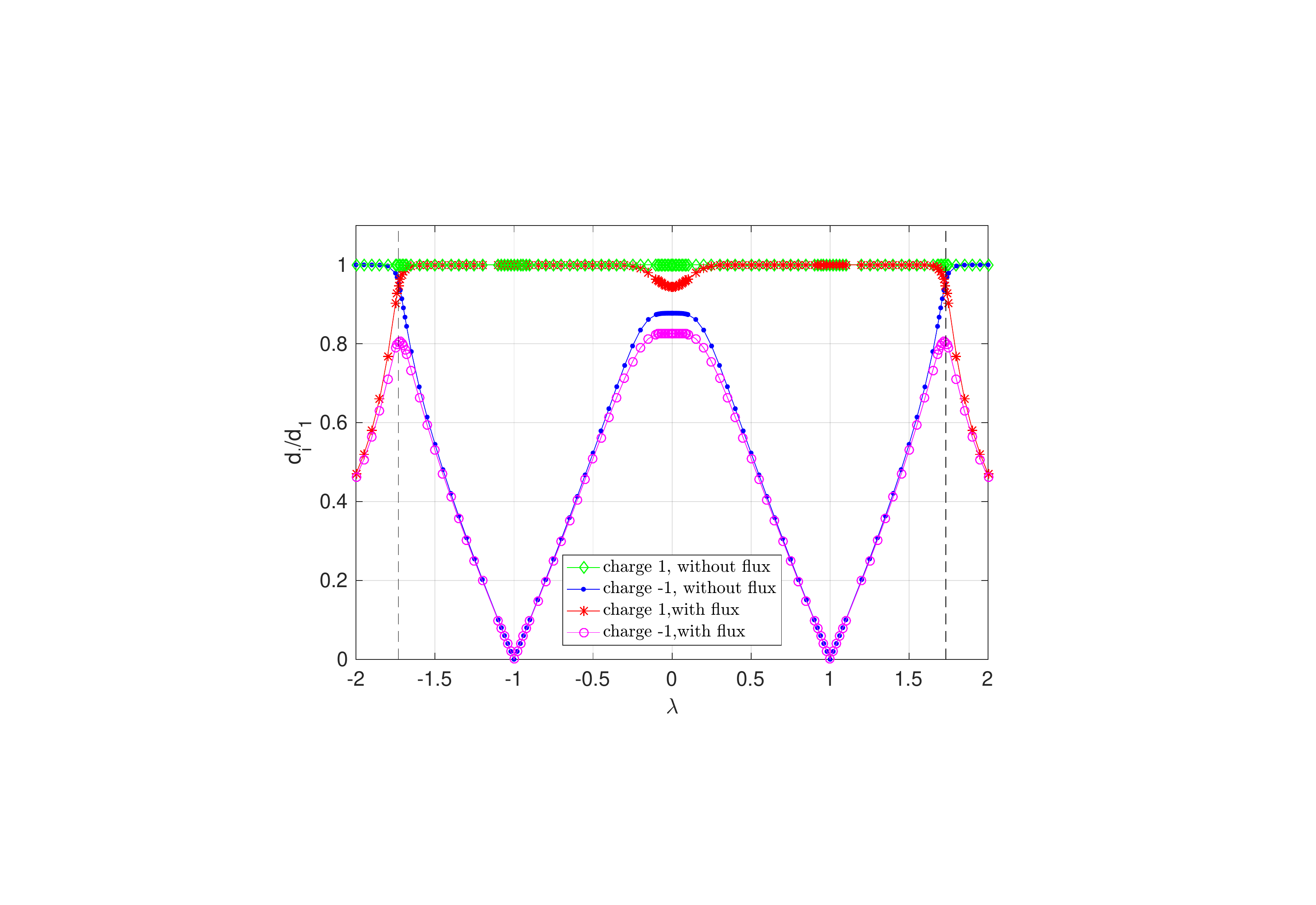}
\caption{The dominant eigenvalues of four different sectors of the transfer
operator. `Charge 1' and `charge -1' are $\mathbb{Z}_{2}$ charge of $U_{%
\protect\phi }\otimes U_{\protect\phi }$. `With flux' or `without flux'
denotes the eigenvalues obtained from $\mathbb{T}$ or $\mathbb{T}_{\protect%
\phi }^{\protect\phi }$.}
\label{Sectors}
\end{figure}

To reveal the mechanism of the quantum topological phase transitions, the
transfer operator spectra are separated into different topological sectors
and the useful information can be extracted from dominant eigenvalues
belonging to each topological sectors. Since both $\mathbb{T}$ and $\mathbb{T%
}_{\phi }^{\phi }$ commute with $U_{\phi }\otimes U_{\phi }$, their
eigenstates carry the $\mathbb{Z}_{2}$ charges of $U_{\phi }\otimes U_{\phi
} $. According to the $\mathbb{Z}_{2}$ charges, the eigenvalues can be
divided into four sectors, corresponding to four types of anyons of $\mathbb{%
Z}_{2}$ topological order \cite{PEPSTO,Shadowofanyons}. In Fig.\ref{Sectors}%
, we display the dominant eigenvalues of the transfer operator $\mathbb{T}%
\oplus \mathbb{T}_{\phi }^{\phi }$ in the different topological sectors as a
function of $\lambda $. In the vicinity of $\lambda \simeq 1.73$, the
dominant eigenvalues of four topological sectors become degenerate in the
thermodynamical limit as $\lambda $ is varied to approach the symmetry
breaking phase. In the symmetry breaking phase, the sectors with the same
flux and different charges are degenerate, suggesting that the full $\mathbb{%
Z}_{2}\times \mathbb{Z}_{2}$ symmetry of the transfer operator is broken and
the bosonic charge condensation occurs. On the other hand, the dominant
eigenvalues of the sectors with the flux insertion are apparently smaller
than those of sectors without flux, signifying the flux excitations must be
confined into pairs \cite%
{Shadowofanyons,AnyonsCondensation,AnyonCondensation2}.

As we expected, the dominant eigenvalues in the vicinity at $\lambda =-1.73$
show the same behavior as that at $\lambda =1.73$. However, the topological
sectors have different interpretations. The anyons in the toric code phase
and double semion phase have the one-to-one correspondence: the electric
charge $\mathbf{e}$ corresponds to the semion-antisemion pair $\mathbf{b}$;
the flux $\mathbf{m}$ corresponds to the semion $\mathbf{s}$, and the
fermion $\mathbf{f}$ correspond to anti-semion $\mathbf{\overline{s}}$.
Therefore, the phase transition at $\lambda =-1.73$ can also be interpreted
as the condensation of $\mathbf{b}$ with the semion $\mathbf{s}$ and
anti-semion $\mathbf{\overline{s}}$ confinement. Nevertheless, the phase
transition at $\lambda =0$ does not belong to such anyon condensation
picture.

\section{Mapping to classical statistical models}

It is known that a class of two-dimensional quantum systems with their
many-body ground state wave functions can be mapped to the partition
functions of the classical statistical systems, because their ground state
wave functions can be written in terms of the Boltzmann weights \cite%
{verstraete-cirac,QuantumDimmerRK,ardonne,Criticality,CQCP,RKmodel}. In
order to explore the topological phase transitions in the above phase
diagram, we would like to derive the corresponding classical statistical
models of the quantum wave functions with $\mathbb{Z}_{2}$ topological order.

\begin{figure}[tbp]
\includegraphics[width=8.5cm]{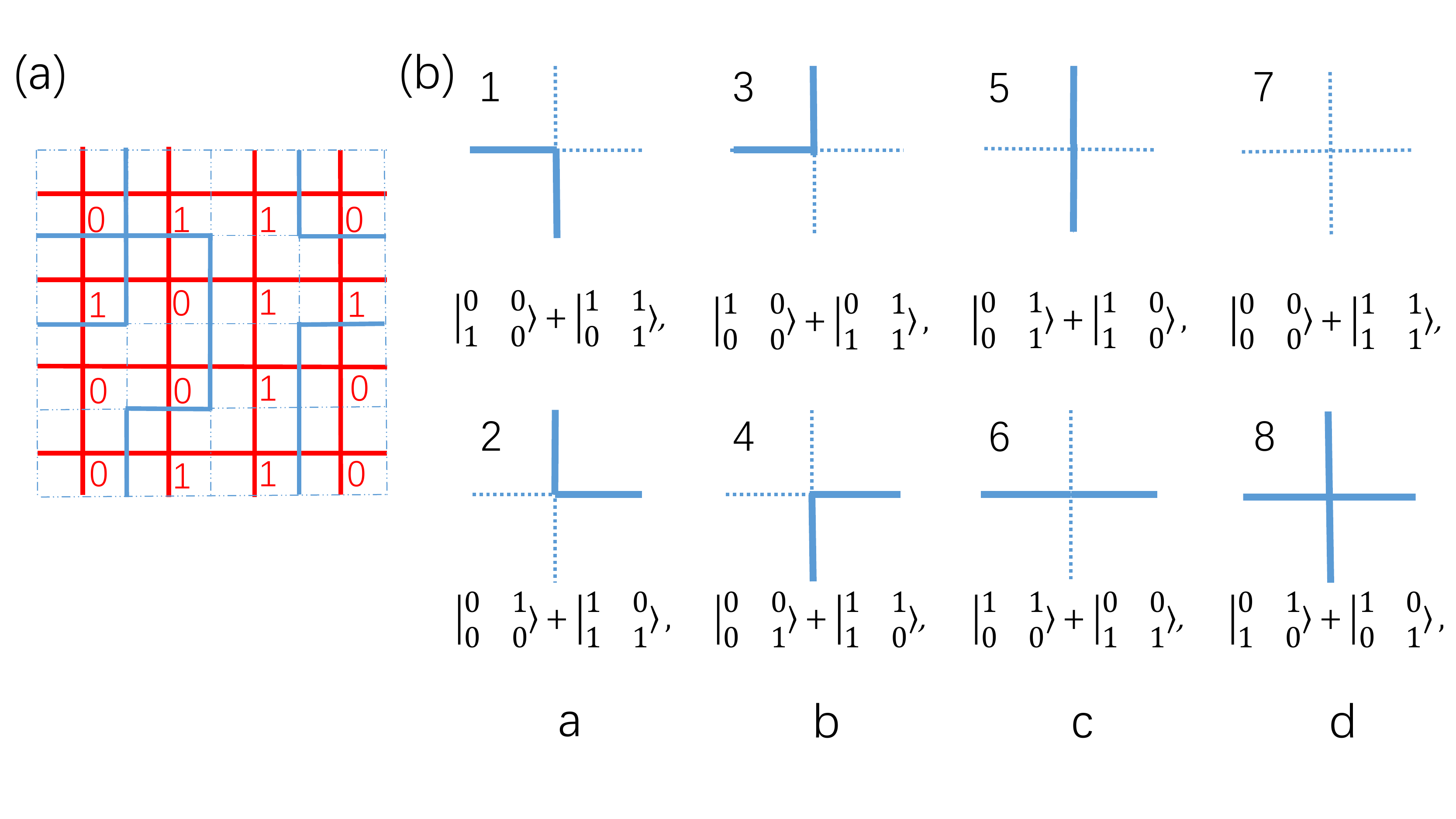}
\caption{(a) Each spin configuration on the red lattice (and its opposite
configuration) is associated to a domain wall configuration on the blue
lattice. (b) The $\mathbb{Z}_{2}$ symmetric superpoistions of the intra-site
spin types have the one-to-one correspondence to the eight-vertex types,
where $a$, $b$, $c$ and $d$ are the Boltzmann weights of the vertex types in
each column, seperately. The number of each vertex type is shown in the
upper left corner. }
\label{EightVertex}
\end{figure}

To map onto a classical statistical model, we have to implement the local tensor 
contractions of the tensor network wave functions. Notice that the virtual degrees 
of freedom of the tensor network states are $\mathbb{Z}_{2}$ spins, and their physical
degrees of freedom are the domain walls. Two spin configurations correspond
to one domain wall configuration, one of them is shown in Fig. \ref%
{EightVertex}(a). A correspondence of the spin types of the intra-site
plaquette and domain wall types of a vertex on the blue lattice can be
figured out, displayed in Fig.\ref{EightVertex}(b). Since the domain wall
configurations of the quantum wave functions can be built by the eight
different kinds of domain walls on the vertices of the blue lattice, the
configurations of the quantum ground states are as the same as the
configurations of the classical eight-vertex model (up to a transformation
on one of the sublattices). There are four states $|0011\rangle $, $%
|1001\rangle $, $|0110\rangle $ and $|1100\rangle $ of the intra-site
plaquettes weighted by $\lambda $, which correspond to the $5$-th and $6$-th
domain wall types on a vertex, so the general tensor network wave functions
can be expressed as
\begin{eqnarray}
|\Psi (\lambda <0)\rangle &=&\sum_{\{i_{e}\}}\lambda ^{n_{5}+n_{6}}\prod
w(s_{\alpha }s_{\beta }s_{\gamma }s_{\delta })|i_{1}i_{2}i_{3}\cdots \rangle
,  \notag \\
|\Psi (\lambda >0)\rangle &=&\sum_{\{i_{e}\}}\lambda
^{n_{5}+n_{6}}|i_{1}i_{2}i_{3}\cdots \rangle ,  \label{TOwavefunction}
\end{eqnarray}%
where $n_{j}$ is the total number of the $j$-type vertices in the closed
domain wall configuration, $w(s_{\alpha }s_{\beta }s_{\gamma }s_{\delta })$
has been defined in Sec. II, and the product runs over all the intra-site
plaquettes of the configurations $\{s_{\alpha }s_{\beta }s_{\gamma
}s_{\delta }\}$ that the configurations $\{i_{e}\}$ correspond to. Although
one domain wall configuration corresponds to the two spin configurations,
the quantity $\prod w(s_{\alpha }s_{\beta }s_{\gamma }s_{\delta })$ is the
same for the two spin configurations.

In the eight-vertex model \cite{ExactSolvableModel}, the Boltzmann weights
of vertex types in the each column of Fig.\ref{EightVertex}(b) are denoted
by $a$, $b$, $c$ and $d$, and the corresponding partition function of the
eight-vertex model is given by
\begin{equation}
\mathcal{Z}=\sum_{\{i_{e}%
\}}a^{n_{1}+n_{2}}b^{n_{3}+n_{4}}c^{n_{5}+n_{6}}d^{n_{7}+n_{8}}.
\label{partition_function_of 8v}
\end{equation}%
The quantum-classical correspondence is usually given by identifying the
norm of the many-body wave function with the partition function:
\begin{equation}
\mathcal{Z}=\langle \Psi (\lambda )|\Psi (\lambda )\rangle
=\sum_{\{i_{e}\}}\lambda ^{2(n_{5}+n_{6})}.
\end{equation}
Compare to Eq. (\ref{partition_function_of 8v}), we immediately notice the
Boltzmann weights
\begin{equation}
a=b=d=1,\text{ }c=\lambda ^{2},
\end{equation}%
which indicate that the phases in the range $\lambda<0$ of the phase diagram
is equivalent to the phases in the range $\lambda>0$. Namely, from the
viewpoint of the statistical models, both double semion and toric code
phases belong to the same paramagnetic disordered phase. Furthermore, it is
well-established that the eight-vertex model becomes critical under the
condition \cite{ExactSolvableModel}: $c=a+b+d$. From the exact solutions of
the eight-vertex model, the exact positions of the QCPs should be at $%
\lambda =\sqrt{3}$, which are perfectly consistent with our numerical
results. In the limit of $c=0$, the eight-vertex model is reduced to the
critical six-vertex model, corresponding to the QCP at $\lambda =0$.

Moreover, it has been known that the effective field theory of the critical
eight-vertex model is described by the two-dimensional Euclidean massless free boson CFT
compactified on a circle with a radius \cite{ardonne}%
\begin{equation}
R^{2}=\frac{4}{\pi }\text{arccot}(\sqrt{cd}),\ \text{for}\ c=d+2,
\end{equation}%
which gives rise to $R=\sqrt{2/3}$. Because the free boson CFT with the
compactified radii $R$ and $2/R$ are dual to each other \cite{CFT}, the the
\textit{dual} CFT should have $R=\sqrt{6}$. The critical exponents of the
correlation length at $\lambda =\sqrt{3}$ are thus known as $\nu =\nu
^{\prime }=3/4$, which are also close to our numerical values.

Although the exact phase transition points can be acquired by mapping the
quantum wave functions to the classical statistical models, the nontrivial
sign factor $\prod w(s_{\alpha }s_{\beta }s_{\gamma }s_{\delta })$ in the
wave functions $|\Psi (\lambda <0)\rangle $ has completely vanished. In other
words, the distinct topological nature between the tensor network states $%
|\Psi (\lambda <0)\rangle $ and $|\Psi (\lambda >0)\rangle $ vanishes, and
both the double semion and toric code wave functions are mapped to the same
classical eight-vertex model. In this sense, the quantum-classical mapping
has \textit{not} kept all the important information on the quantum many-body
phases, so the quantum topological phase transition between two distinct
topologically ordered phases at $\lambda =0$ can not be understood from the
critical six-vertex model. More importantly, the ground states of a
topologically ordered phase have a degeneracy on a torus, and different
ground states can be related by inserting the MPOs into the tensor network
states, which is not included in the quantum-classical mapping as well. So
we have to decode the critical properties of the QCPs by employing other
methods in the tensor network approach.

\section{Conformal quantum criticalities}

\subsection{Central charges}

Although the phase transitions from the topological phases to the symmetry
breaking phase can be explained by anyon condensation \cite%
{AnyonsCondensation,AnyonCondensation2}, the quantum phase transition
between two topological phases is still mysterious. Meanwhile only a few
attempts have considered the phase transitions between the toric code and double
semion models \cite{TC-DS,AnyonsCondensation,AnyonCondensation2}. In this
section, we will study the QCPs from the full spectra of the transfer
operators without/with the flux insertions, and the underlying field
theories of the QCPs can be revealed.

\begin{figure}[tbp]
\includegraphics[width=9cm, trim=150 120 100
150,clip]{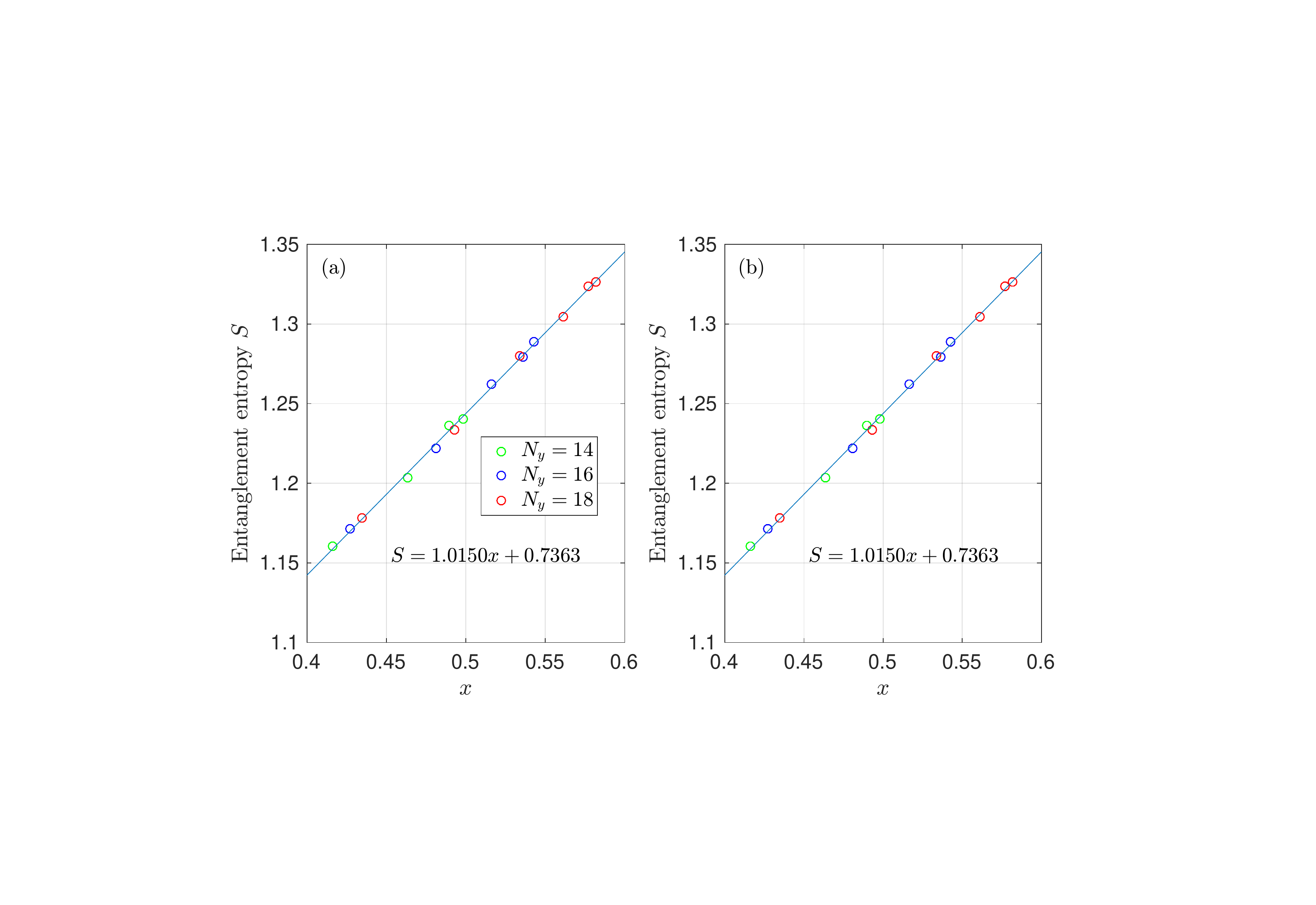}
\caption{(a) The entanglement entropy is calculated from the dominant
eigenstate of the transfer operator $\mathbb{T}(\protect\lambda =0)$ and
fitted with the Calabrese-Cardy formula to extract the central charge. Here $%
x(n,N_{y})=\frac{1}{3}\text{ln}[\frac{N_{y}}{\protect\pi }\text{sin}(\frac{n%
\protect\pi }{N_{y}})]$ and $n$ is the lattice site number of the subsystem
in the bipartition. (b) The similar analysis for the transfer operator $%
\mathbb{T}(\protect\lambda =\pm \protect\sqrt{3})$.}
\label{QCP1}
\end{figure}

It has been demonstrated that the spectrum of the transfer operator defined
from the wave function norm contains much useful information of the bulk
properties of the tensor network states\cite{TransferMatrices,Shadowofanyons}%
. We have used the dominant eigenvalues of the transfer operator to deduce
the correlation length. The previous numerical correlation length shows that
$\xi \propto N_{y}$ at three QCPs, so we can further calculate the
entanglement entropy from the corresponding dominant eigenvector of the
transfer operator $\mathbb{T}$. After the numerical entanglement entropy is
obtained, we can fit the numerical results with the Calabrese-Cardy formula
\cite{CCformula},
\begin{equation}
S(n,N_{y})=\frac{c}{3}\ln \left[ \frac{N_{y}}{\pi }\sin \left( \frac{\pi n}{%
N_{y}}\right) \right] +\text{const.},
\end{equation}%
where $n$ is the segment length chosen in the bipartition of the dominant
eigenvector with a given circumference $N_{y}$, and the central charge $c$
for the QCPs are extracted and shown in Fig.\ref{QCP1}(a) and (b). The
fitting central charge is estimated as $c\approx 1$ for both $\lambda =0$
and $\lambda =\pm \sqrt{3}$, indicating that the quantum criticalities of all
these three QCPs are characterized by a two-dimensional compactified free
boson CFT!

It has been argued that the moduli of eigenvalues of the transfer operator
correspond to the minimum of the excitation spectrum of the bulk system \cite%
{TransferMatrices}, and the one-dimensional quantum transfer operator is a
manifestation of the holographic bulk-boundary correspondence \cite{ESofPEPS}%
. Instead of solving the model Hamiltonians in two dimension, it is more
efficient to extract the properties of low-energy excitations from the
transfer operator spectra. When the quantities $\epsilon _{i}=-$ln$|\frac{%
d_{i}}{d_{1}}|$ defined from the eigenvalues of the transfer operators are
carefully analyzed, we surprisingly find that the finite-size scaling law $%
\epsilon _{i}\propto \frac{1}{N_{y}}$ exactly satisfies at the critical
points, suggesting that $\epsilon _{i}$ can be regarded as the spectral
levels of the finite-size spectrum of the corresponding CFTs.

\subsection{Finite-size spectra at $\protect\lambda =\protect\sqrt{3}$}

The finite-size spectrum $\epsilon _{i}$ at $\lambda =\sqrt{3}$ can be a
function of the lattice momentum, which is extracted from the translation
symmetry. Since the transfer operator $\mathbb{T}\oplus \mathbb{T}_{\phi
}^{\phi }$ has the MPO symmetry of $U_{\phi }\otimes U_{\phi }$, the
eigenstates of this transfer operator can be decomposed into two different
sectors by the $\mathbb{Z}_{2}$ charges of $U_{\phi }\otimes U_{\phi }$.
Moreover, the transfer operator $\mathbb{T}$ has the translation symmetry $%
T\otimes T$, where $T$ separately acts on the bra and ket layers and shifts
the $j$-th half plaquette to the $(j-1)$-th half plaquette. Although the
transfer operator $\mathbb{T}_{\phi }^{\phi }$ breaks the translation
symmetry, a modified translation symmetry \cite%
{ABEffect,Wang-Santos-Wen,SymmetryDefect} can be found as $\tilde{T}%
=TX_{j}X_{j+1}$ for $\lambda >0$ and $\tilde{T}=TCZ_{j,j+1}X_{j}X_{j+1}$ for
$\lambda <0$, where the flux is inserted on the $j$-th half plaquettes, see
Fig. \ref{TO-flux}(a), and the modified translation operators acts on the $j$%
-th and $(j+1)$-th plaquettes.
\begin{figure}[tbp]
\includegraphics[width=9cm]{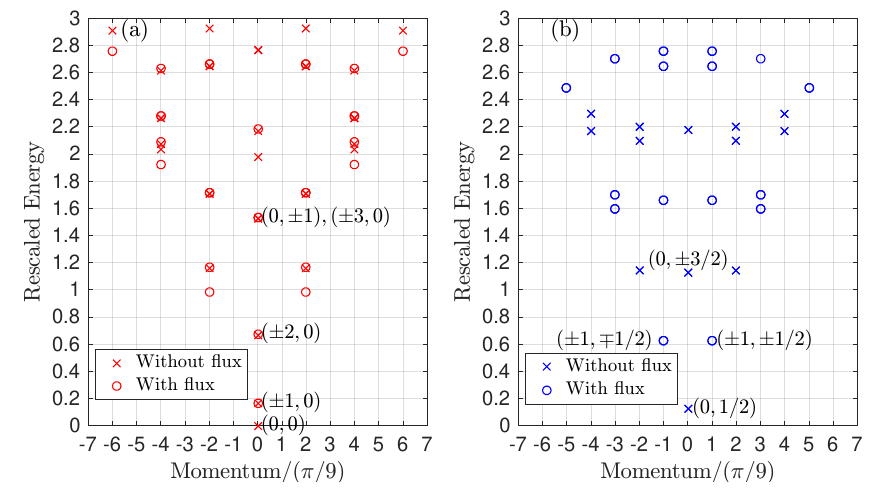}
\caption{(a) The spectrum of the charge $1$ sector of the transfer operator $%
\mathbb{T}\oplus \mathbb{T}_{\protect\phi }^{\protect\phi }$ at $\protect%
\lambda =\protect\sqrt{3}$. (b) The corresponding charge $-1$ sector. Here
the momentum is calculated from the squared translation operator $T_{l}^{2}$
or $\tilde{T}_{l}^{2}$ with $N_{y}=18$ lattice sites and the spectral levels
are rescaled such that they can perfectly match the scaling dimensions. The
red and blue signs marks the $\pm 1$ charges of the symmetry $U_{\protect%
\phi }\otimes U_{\protect\phi }$. The cross and circle markers the levels
from $\mathbb{T}$ or $\mathbb{T}_{\protect\phi }^{\protect\phi }$. The
scaling dimensions of the primary fields are denoted by $(e,m)$. }
\label{QCP2Spectrum}
\end{figure}

However, the eigenvalue spectra $\epsilon _{i}$ of $\mathbb{T}$ and $\mathbb{%
T}_{\phi }^{\phi }$ do not yield the complete CFT spectra directly. The
complete CFT spectrum includes both the $\mathbb{Z}_{2}$ charge $1$ and $-1$
sectors of the operator $\mathbb{T}\oplus \mathbb{T}_{\phi }^{\phi }$. Figs.%
\ref{QCP2Spectrum}(a) and (b) show the finite-size spectra $\epsilon _{i}$
with respect to the momenta, and the spectral levels are rescaled such that
the $\epsilon _{i}$ can be fitted to the scaling dimensions of the primary
fields of the compactified free boson CFT:
\begin{equation}
\Delta (e,m)=\frac{e^{2}}{R^{2}}+\frac{m^{2}R^{2}}{4},
\end{equation}%
where $e$ denotes the total angular momenta of the primary fields, $m$ is
the winding numbers of the primary fields, $s=em$ are the conformal spins of
primary fields \cite{CFT}, and the compactified radius has been found as $R=%
\sqrt{6}$. Both quantum numbers of $e$ and $m$ are normally chosen as
integers in the charge $1$ sector, corresponding to the CFT spectrum of the
critical eight-vertex model. From the scaling dimensions $\Delta =h+\bar{h}$
and conformal spins $s=h-\bar{h}$, we can determine the primary fields with
the conformal dimensions $(h,\bar{h})$, see the Table I. The reason why the
sector with the charge $1$ corresponds to the finite-size spectrum of the
eight-vertex model is that this sector corresponds to the ground-state wave
function in which the domain walls are closed. It is also known that the
compactified free boson CFT with $R=\sqrt{k}$ is denoted as the $U(1)_{k}$
CFT \cite{CFT_on_torus}, and the charge $1$ sector thus belongs to the $%
U(1)_{6}$ CFT.
\begin{table}[h]
\caption{Primary fields, scaling dimensions, and conformal spins of the
charge $1$ sector for the QCP at $\protect\lambda =\protect\sqrt{3}$.}
\label{CFT-sqrt3}%
\begin{tabular}{cccc}
\hline\hline
\multicolumn{2}{c}{Primary fields} & {Scaling dimensions} & {Conformal spins}
\\
$(h,\bar{h})$ & $(e,m)$ & $\Delta=h+\bar{h}$ & $s=h-\bar{h}$ \\
\colrule $(0,0)$ & $(0,0)$ & $0$ & $0$ \\
$(3/4,3/4)$ & $(0,\pm1)$ & $3/2$ & $0$ \\
$(1,0)$ & / & $1$ & $1$ \\
$(0,1)$ & / & $1$ & $-1$ \\
$(1/3,1/3)$ & $(\pm2,0)$ & $2/3$ & $0$ \\
$(1/12,1/12)$ & $(\pm1,0)$ & $1/6$ & $0$ \\
\botrule &  &  &
\end{tabular}%
\end{table}

However, in the charge $-1$ sector, the conformal dimensions of primary
fields are independent of $R$ and are fixed as $1/16$ and $9/16$ (Ref. \cite%
{AppliedCFT}), so the scaling dimensions can be expressed as $\Delta(e,m)=%
\frac{e^2}{2}+\frac{m^2}{2}$. The primary fields, scaling dimensions and
their conformal spins are given in Table II. Moreover, the momenta of
spectral levels are determined as $\frac{2\pi }{N_{y}}(\mathbb{Z}+\frac{1}{2}%
)$, representing the topological anyon excitations. Therefore, the quantum
criticality of this QCP exhibits two different characteristics: the
spontaneous symmetry breaking and confinement of the topological anyon
excitations.
\begin{table}[h]
\caption{Primary fields, scaling dimensions, and conformal spins of the
charge $-1$ sector for the QCP at $\protect\lambda =\protect\sqrt{3}$.}%
\begin{tabular}{cccc}
\hline\hline
\multicolumn{2}{c}{Primary fields} & {Scaling dimensions} & {Conformal spins}
\\
$(h,\bar{h})$ & $(e,m)$ & $\Delta=h+\bar{h}$ & $s=h-\bar{h}$ \\
\colrule $(1/16,1/16)$ & $(0,\pm1/2)$ & $1/8$ & $0$ \\
$(1/16,9/16)$ & $(\pm1,\mp1/2)$ & $5/8$ & $-1/2$ \\
$(9/16,1/16)$ & $(\pm1,\pm1/2)$ & $5/8$ & $1/2$ \\
$(9/16,9/16)$ & $(0,\pm3/2)$ & $9/8$ & $0$ \\
\botrule &  &  &
\end{tabular}%
\end{table}

Actually the combination of the charge $1$ and charge $-1$ sectors can be
interpreted as the $\mathbb{Z}_{2}$ orbifold $U(1)_{6} $ free boson CFT with
the compactified radius $R=\sqrt{6}$. Fig. \ref{QCP2Spectrum}(a) corresponds
to the spectrum of the untwisted sector, while Fig. \ref{QCP2Spectrum}(b)
represents the spectrum of the twisted sector \cite{AppliedCFT,Curiosities}.
Since the $\mathbb{Z}_{2}$ orbifold CFT characterizes the critical
properties of the Ashkin-Teller model, the present QCP belongs to the
universality class of the Ashkin-Teller model at the $\mathbb{Z}_{4}$
parafermion point \cite{AppliedCFT,Li_Yang_Tu_Cheng2015}.

\subsection{Finite-size spectra at $\protect\lambda =-\protect\sqrt{3}$}

As we discussed in the previous section, the QCP at $\lambda =-\sqrt{3}$ is
similar to the QCP at $\lambda =\sqrt{3}$, and the spectra of the transfer
operators are as the same as Fig. \ref{QCP2Spectrum}. So the QCP also
belongs to the universality class of the Ashkin-Teller model at the $\mathbb{%
Z}_{4}$ parafermion point, and the quantum phase transition is caused by the
boson condensation and semion/anti-semion confinement. Therefore, just from
the transfer operator spectra, we can not make the distinction between the
QCP at $\lambda =-\sqrt{3}$ and the QCP at $\lambda =\sqrt{3}$. However, in
the entanglement spectrum of the double semion phase, we have known that the
semion and anti-semion sectors should carry quarter-integer momenta $\frac{%
2\pi }{N_{y}}(n\pm \frac{1}{4})$ with $n\in \mathbb{Z}$. Then it is expected
that the spectral levels of the transfer operators in the topological
sectors with charge $-1$ should also carry quarter integer momenta in the
double semion phase. Nevertheless, the eigenvalues of the lattice
translational operator $T$ are given by $\frac{2\pi }{N_{y}}(n\pm \frac{1}{4}%
)$, and thus the eigenvalues of $T\otimes T$ are obtained as either $\frac{%
2\pi }{N_{y}}(n+\frac{1}{2})$ or $\frac{2\pi }{N_{y}}n$. Unlike the toric
code phase, the spectral momenta (conformal spins) in the semion and
anti-semion sectors are not properly associated to the topological spins.
Distinguishing these two critical points can be seen in the wave functions
of the toric code phase and double semion phase. The essential difference
just exists in the quasi-particle braiding statistics, which can be obtained
from the modular matrices. So it is reasonable that we can not distinguish
them via the transfer operator spectra and the corresponding static
correlators. We have resolved this problem by investigating the dominant
eigenvectors of the transfer operators and some arguments are provided in
Appendix \ref{Momentum_polorization}.

\subsection{Finite-size spectra at $\protect\lambda =0$}

The QCP at $\lambda =0$ is very exotic, describing the continuous
topological phase transition between two topologically ordered phases.
Essentially different from the QCPs at $\lambda =\pm \sqrt{3}$, it is more
natural to classify the finite-size spectra of the transfer operators at $%
\lambda =0$ according to the absence or presence of the flux insertion. As
we discussed in Sec.III, we can have four different transfer operators: $%
\mathbb{T}$, $\mathbb{T}_{\text{cz}}$, $\mathbb{T}_{\text{x}}^{\text{x}}$,
and $\mathbb{T}_{\text{czx}}^{\text{x}}$, leading to topologically different
sectors. The corresponding eigenstates of $\mathbb{T}_{\text{x}}^{\text{x}}$%
, $\mathbb{T}_{\text{czx}}^{\text{x}}$ and $\mathbb{T}_{\text{cz}}$ carry
the symmetry charges of $U_{\text{x}}\otimes U_{\text{x}}$, $U_{\text{x}%
}\otimes U_{\text{czx}}$ and $\mathbbm{1}\otimes U_{\text{cz}}$,
respectively. Since $\mathbb{T}$ commutes with $U_{\text{x}}\otimes U_{\text{%
x}}$, $U_{\text{cz}}\otimes U_{\text{cz}}$ and $U_{\text{czx}}\otimes U_{%
\text{czx}}$ and the eigenstates preserve these symmetries, we can use the
quantum numbers associated with these symmetries to label the spectral
levels.
\begin{figure*}[tbp]
\includegraphics[width=18cm]{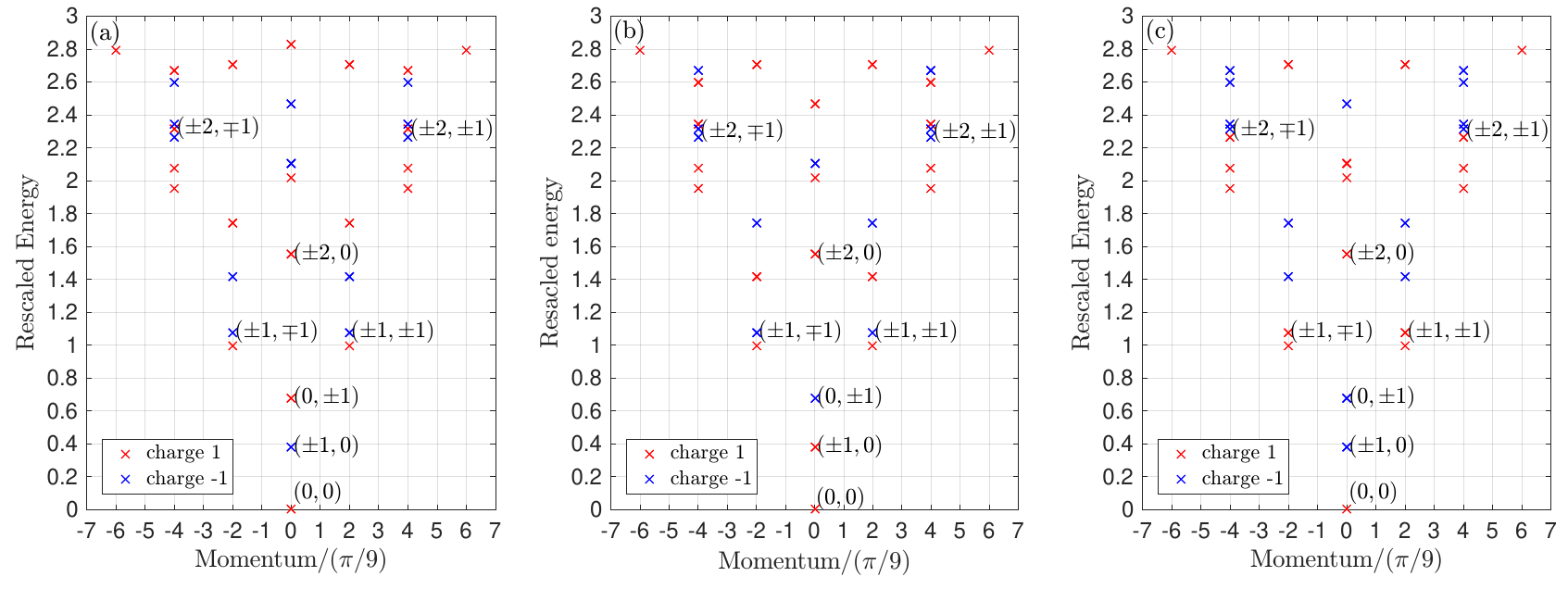}
\caption{The spectra of the transfer operator $\mathbb{T}$ for the QCP at $%
\protect\lambda =0$, where the spectral levels are labelled by $\mathbb{Z}%
_{2}$ symmetry charge of $U_{\text{x}}\otimes U_{\text{x}}$ in (a), the $%
\mathbb{Z}_{2}$ symmetry charge of $\mathbbm{1}\otimes U_{\text{cz}}$ in
(b), and the $\mathbb{Z}_{2}$ symmetry charge of $U_{\text{x}}\otimes U_{%
\text{czx}}$ in (c). Here the transfer operators include $N_y=18$ lattice
sites and the spectral levels are rescaled such that they perfectly match
the scaling dimensions. The red and blue signs denote the $\mathbb{Z}_{2}$
charge $\pm 1$ of the corresponding symmetry. The primary fields are
expressed by $(e,m)$.}
\label{QCP3Spectrum}
\end{figure*}

Moreover, the insertion of MPOs breaks the translation symmetry $T\otimes T$%
, so we need to use the modified translational operators \cite%
{ABEffect,Wang-Santos-Wen,SymmetryDefect}. For $\mathbb{T}_{\text{x}}^{\text{%
x}}$, the modified translation symmetry is expressed as $(TX_{j}X_{j+1})%
\otimes (TX_{j}X_{j+1})$, where the MPOs are inserted on the plaquettes of
the $j $-th row and the operators act on the vicinity edges of the $j$ and $%
(j+1)$-th half plaquette. The modified translational symmetry for $\mathbb{T}%
_{\text{cz}}$ is given by $T\otimes (TCZ_{j,j+1})$, while the modified
translational operator for $\mathbb{T}_{\text{czx}}^{\text{x}}$ is $%
(TX_{j}X_{j+1})\otimes (TCZ_{j,j+1}X_{j}X_{j+1})$.

In Fig. \ref{QCP3Spectrum}(a), the spectrum of $\mathbb{T}$ is displayed,
and the spectral levels are labelled by the quantum numbers of the
translation symmetry $T\otimes T$ and $\mathbb{Z}_{2}$ charge of $U_{\text{x}%
}\otimes U_{\text{x}}$. By carefully fitting the spectra with the scaling
dimensions $\Delta (e,m)=\frac{e^{2}}{R^{2}}+\frac{m^{2}R^{2}}{4}$, we found
that the full spectrum can be described by the $U(1)$ compactified free
boson CFT with a compactified radius $R\approx \sqrt{8/3}$. The conformal
dimensions of primary fields for such a theory are determined by
\begin{equation}
h=\left (\frac{e}{R}+\frac{mR}{2}\right )^{2}\text{, }\bar{h}=\left (\frac{e%
}{R}-\frac{mR}{2}\right )^{2}.
\end{equation}%
Moreover, the symmetry charge of the $U_{\text{x}}\otimes U_{\text{x}}$
carried by the spectral levels satisfies the rule $(-1)^{e}$, and the fields
in the same conformal tower have the identical $\mathbb{Z}_{2}$ charge. Fig. %
\ref{QCP3Spectrum}(b) is also the spectrum of $\mathbb{T}$ but the spectral
levels are labelled by the $\mathbb{Z}_{2}$ charge of the symmetry $%
\mathbbm{1}\otimes U_{\text{cz}}$, which is determined by the rule $(-1)^{m}
$. When the spectrum of $\mathbb{T}$ is further labelled by the charge of
the symmetry $U_{\text{x}}\otimes U_{\text{czx}}$ in Fig. \ref{QCP3Spectrum}%
(c), the low-energy fields satisfy the rule $(-1)^{e+m}$, similar to the
entanglement spectrum of the $\mathbb{Z}_{2}$ nontrivial SPT phase \cite%
{ABEffect,SCofSPT,ChiralSymmetry}. These features are summarized in the
first two columns of Table. \ref{rules}. 

\begin{figure*}[tbp]
\includegraphics[width=18cm]{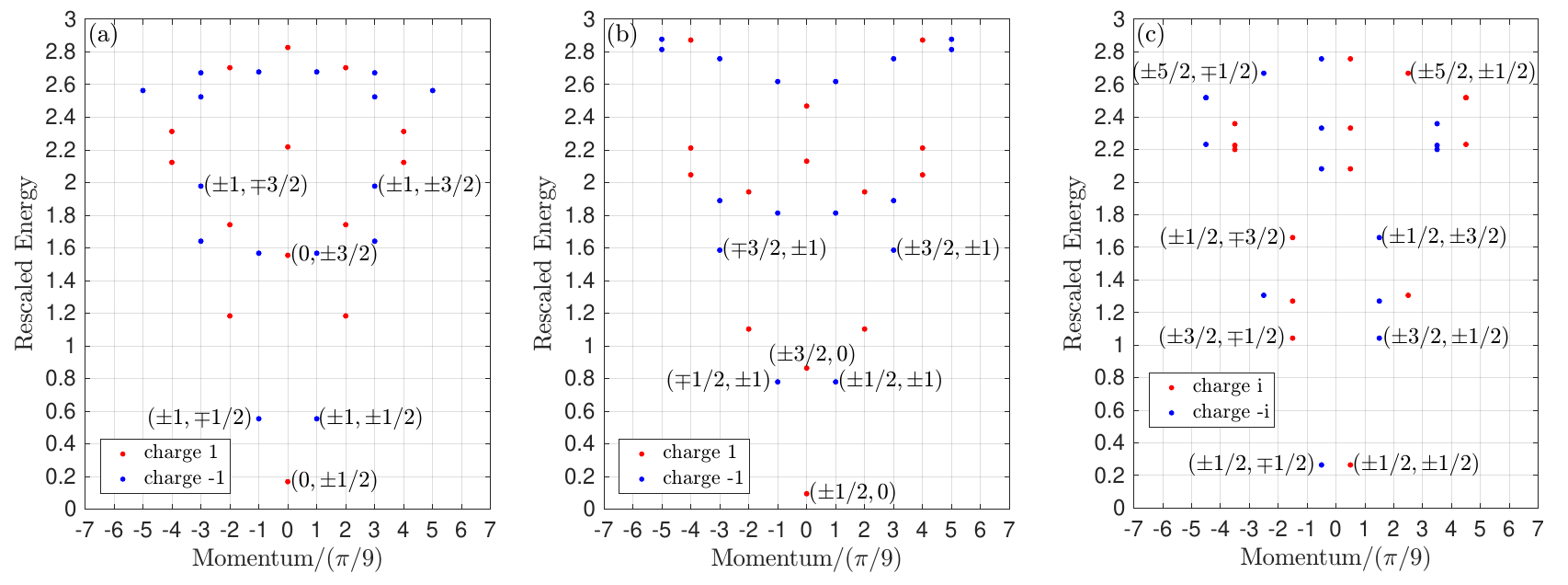}
\caption{The spectra of the transfer operators with the flux insertions. (a)
$\mathbb{T}_{\text{x}}^{\text{x}}$, (b) $\mathbb{T}_{\text{cz}}$ and (c) $%
\mathbb{T}_{\text{czx}}^{\text{x}}$. The red and blue dots stand for the $%
\mathbb{Z}_{2}$ charges $\pm 1$ of the corresponding symmetry. Here the
transfer operators include $N_{y}=18$ lattice sites and the momentum is
calculated from the square of the modified translation operator and the
energy levels are rescaled such that they perfectly match the scaling
dimensions of the primary fields $(e,m)$. }
\label{QCP1Spectrum}
\end{figure*}

Furthermore, we can also calculate the finite-size spectra of the transfer
operators with the flux insertions, describing the topological excitations
with fractionalized quantum numbers. The resulting spectra are also
well-fitted into the $U(1)$ compactified free boson CFT with the
compactified radius $R=\sqrt{8/3}$, and can be divided into three sets of
spectra. Fig. \ref{QCP1Spectrum}(a) shows the first set of spectra from the
operator $\mathbb{T}_{\text{x}}^{\text{x}}$ with the $\mathbb{Z}_{2}$ charge
of $U_{\text{x}}\otimes U_{\text{x}}$, and the winding numbers $m$ are
determined as half-integers and the angular momenta of boson fields $e$ as
integers. The $\mathbb{Z}_{2}$ charges are given by $(-1)^{e}$, the levels
in the same conformal tower have the identical $\mathbb{Z}_{2}$ charge, but
the momenta of the spectral levels with the charge $-1$ are half-integers $%
\frac{2\pi }{9}(\mathbb{Z}+\frac{1}{2})$. The second set of spectra from the
operator $\mathbb{T}_{\text{cz}}$ is displayed in Fig. \ref{QCP1Spectrum}%
(b), where the eigenstates of $\mathbb{T}_{\text{cz}}$ carry the charge of $%
\mathbbm{1}\otimes U_{\text{cz}}$ and the scaling dimensions of the primary
fields are given by the integer $m$ and half-integers $e$. But the symmetry
charges are determined by $(-1)^{m}$, and the levels with the charge $-1$
also carry half-integer momenta. Fig. \ref{QCP1Spectrum}(c) shows the third
set of spectra from the operator $\mathbb{T}_{\text{czx}}^{\text{x}}$, where
the eigenstates carry the charges of $U_{\text{x}}\otimes U_{\text{czx}}$
and both the quantum numbers $m$ and $e$ are half-integers. However, it
should be emphasized that the eigenvalues of $U_{\text{x}}\otimes U_{\text{%
czx}} $ are $\pm i$, which are determined by $-i(-1)^{e+m} $, so that the
the spectral levels carry the \textit{quarter}-integers momenta $\frac{2\pi
}{9}(\mathbb{Z}\pm \frac{1}{4})$. Such a compactified free boson CFT has
been predicted for the entanglement spectrum of the $\mathbb{Z}_{2}$
non-trivial SPT phases with symmetry defects \cite%
{ABEffect,SymmetryDefect,Wang-Santos-Wen,SCofSPT,SPTorbifolds}. These
results are displayed in the last four columns of Table. \ref{rules}.
All these three sets of CFT spectra reflect the rich 
structure of the low-energy excitations at the QCP.

\begin{table}[tbp]
\caption{Symmetry of MPO, symmetry charges of $\mathbb{T}$, $\mathbb{T}_{%
\text{x}}^\text{x}$, $\mathbb{T}_{\text{cz}}$ and $\mathbb{T}_{\text{czx}}^%
\text{x}$, and fractionalization of $e$ and $m$ for the QCP at $\protect%
\lambda =0$. The first two columns show properties of $\mathbb{T}$, the last
four columns display properties of $\mathbb{T}_{\text{x}}^\text{x}$, $%
\mathbb{T}_{\text{cz}}$ and $\mathbb{T}_{\text{czx}}^\text{x}$.}
\label{rules}%
\begin{tabular}{ccccccc}
\toprule {Symmetry} & {Symmetry} & {After S} & {$e$} & {$m$} & {Symmetry} &
\\
{of $\mathbb{T}$} & {charges of $\mathbb{T}$} & {\ transform} &  &  & {%
charges } &  \\
\colrule $U_\text{x}\otimes U_\text{x}$ & $(-1)^e$ & $\mathbb{T}_{\text{x}}^%
\text{x}$ & $\mathbb{Z}$ & $\mathbb{Z}+\frac{1}{2}$ & $(-1)^e$ &  \\
$\mathbbm{1}\otimes U_\text{cz}$ & $(-1)^m$ & $\mathbb{T}_{\text{cz}}$ & $%
\mathbb{Z}+\frac{1}{2}$ & $\mathbb{Z}$ & $(-1)^m$ &  \\
$U_\text{x}\otimes U_\text{czx}$ & $(-1)^{e+m}$ & $\mathbb{T}_{\text{czx}}^%
\text{x}$ & $\mathbb{Z}+\frac{1}{2}$ & $\mathbb{Z}+\frac{1}{2}$ & $%
-i(-1)^{e+m}$ &  \\
\botrule &  &  &  &  &  &
\end{tabular}%
\end{table}

Actually there are close relations between the above two types of spectra.
Considering the fact that the actions of symmetries $U_{\text{x}}\otimes U_{%
\text{x}}$, $\mathbbm{1}\otimes U_{\text{x}}$ and $U_{\text{x}}\otimes U_{%
\text{czx}}$ on the low-energy excitations are given by $(-1)^{e}$, $(-1)^{m}
$ and $-i(-1)^{e+m}$, respectively, we can analytically derive the
corresponding the partition function of compactified boson CFT modular after
the modular $S$ transformation using the Poisson resumation \cite%
{CFT,CFT_on_torus,SCofSPT}. Then the quantum numbers of $e$ and $m$ are
fractionalized into half-integers in different way according to the symmetry
actions for the three cases, as shown in each row of Table. \ref{rules}.
These results coincide with our numerical spectra. Thus, the three sets of
excitations in Fig.\ref{QCP3Spectrum} and those three sets in Fig.\ref%
{QCP1Spectrum} have an exact one-to-one mapping through the modular $S$
transformation.

\section{Discussion and conclusion}

It is intuitively expected that the quantum criticalities of a
two-dimensional quantum system should be described by (2+1)-dimensional
CFTs. However, in our tensor network state approach, the quantum topological
phase transitions of the two-dimensional quantum system are related to
(2+0)-dimensional time-independent CFTs in the sense that the ground state
static correction functions of 2-dimensional quantum systems are equal to
the correlators of (2+0)-dimensional CFTs. How to understand such numerical
results?

In the literature, there exist two different scenarios to approach the
quantum topological phase transitions\cite{castelnovo}. In the first
``Hamiltonian deformation'' approach, where two distinct fixed-point model
Hamiltonians are interpolated with a parameter, a QCP may reach at a
critical value of the parameter and the corresponding effective action has
the Lorentz invariance with the dynamical critical exponent $z=1$, and the
QCP is thus described by a (2+1)-dimensional CFT. In the second approach,
however, the QCP sits on the different path called \textquotedblleft wave
function deformation\textquotedblright in the parameter space, where the
effective action is not Lorentz-invariant, characterized by the dynamical
critical exponent $z>1$ due to the intrinsic space-time asymmetry. In
addition, the corresponding tensor network states with \textit{algebraically}
decaying correlation functions keep a finite bond dimension even at the
critical point \cite{Perturbative_Tensor_Networks}. So a phenomenon of
holographic dimensionality reduction occurs, and the essential information
about the low-energy excitations has been encoded in the one-dimensional
quantum transfer operators without/with the flux insertions. Such QCPs
belong to the so-called Rokhsar-Kivelson type\cite{castelnovo_quantum_2005}
conformal QCPs\cite{ardonne}, because the action is invariant under
conformal transformation of two-dimensional space.

Although the static correlation functions at these QCPs are characterized by
the compactified free boson CFT, the dynamics of these QCPs can be different,
depending on the dynamic symmetry class \cite{CQCP}. Because there is an $U(1)$
symmetry along the critical line, the quantum six-vertex model obeys the scaling
with a dynamical exponent $z=2$. So the QCP at $\lambda=0$ should belong to the
quantum Lifshitz theory \cite{ardonne}. However, the QCPs at $\lambda=\pm\sqrt{3}$
are more difficult to have a concrete conclusion, as the quantum eight-vertex model
possesses dynamical exponent $z\geq2$ in their critical regime, it was suggested
that such conformal QCPs may not be the final stabilized fixed points under the
renormalization group transformation and they may flow to the stable
(2+1)-dimensional Lorentz-invariant QCPs with $z=1$ relativistic dynamics in the
end \cite{ardonne,CQCP}. On the other hand, up to one-loop approximation, a recent 
renormalization group study around this critical point has shown that the dynamical 
exponent does not flow \cite{Dynamical_Lifshitz_theory} in contrast to the
previous classical Monte Carlo study \cite{CQCP}. Thus, further investigations are
certainly needed to clarify this issue.

More importantly, the QCPs at $\lambda =\pm \sqrt{3}$ have two different aspects.
They represent a quantum phase transition from the quantum symmetric phase
to the symmetry breaking phase, corresponding to the conventional
disorder-order phase transition of the eight-vertex model. On the other
hand, they also characterize a quantum topological phase transition from the
topologically ordered (anyon deconfined) phase to the topologically trivial
(anyon confined) phase, which can be explained by the mechanism of anyon
condensation. However, the QCP at $\lambda =0$ \textit{can not} be
understood within such a mechanism, because the quantum phase transition
occurs at the end point of disordered phase of the classical statistical
model. The corresponding criticality contains a rich CFT structure of
low-energy topological excitations.

To summarize, we have proposed the tensor network state approach to the
quantum topological phase transitions and their criticalities in two
dimensions. By gauging a tensor network state of $\mathbb{Z}_{2}$ SPT phases
with a tuning parameter $\lambda $, we have constructed a general tensor
network state wave function for the intrinsically $\mathbb{Z}_{2}$
topological phases, which incorporates the toric code phase, double semion
phase, and the symmetry breaking phase. From the calculation of the
correlation length defined from the one-dimensional quantum transfer
operator, we have mapped out the full phase diagram and identified three
QCPs at $\lambda =\pm \sqrt{3}$ and $\lambda =0$, respectively. Then we have
further proved that the quantum criticalities at these three QCPs can be
extracted from the complete spectra of the transfer operators without/with
the flux insertions. The static correlators of resulting conformal QCPs
should be described by the (2+0)-dimensional time-independent $U(1)$
compactified free boson CFTs. There are many open questions, e.g, how to
expand the present tensor network approach to the topological phase
transitions among $\mathbb{Z}_{3}$ topologically ordered phases? In
addition, the conformal QCPs for the topological phase transitions of the
non-abelian topological phases are also interesting. These issues are left
for future works.

\textit{Acknowledgment.- }The authors would like to thank Guo-Yi Zhu for his
stimulating discussions and acknowledges the support of National Key
Research and Development Program of China (2017YFA0302902).

\appendix

\section{Phase diagram for $Z_{2}$ SPT phases}

\label{Phase_diagram_of_SPT}

As we mentioned before, the transfer operator for SPT states is $\mathbb{T}%
_{0}$, so we can calculate the ground state phase diagram from it. For
larger $\lambda $, the numerical calculation shows that the dominant
eigenvalues of the transfer operator have nearly four-fold degeneracy,
because the splitting between these nearly degenerate eigenvalues becomes
exponentially small when increasing the system size. This observation is
consistent with the previous analysis that the spontaneous symmetry breaking
occurs for larger $|\lambda |$. While for smaller $|\lambda |$, the dominant
eigenvalue is non-degenerate.

\begin{figure}[tbp]
\includegraphics[width=8.5cm]{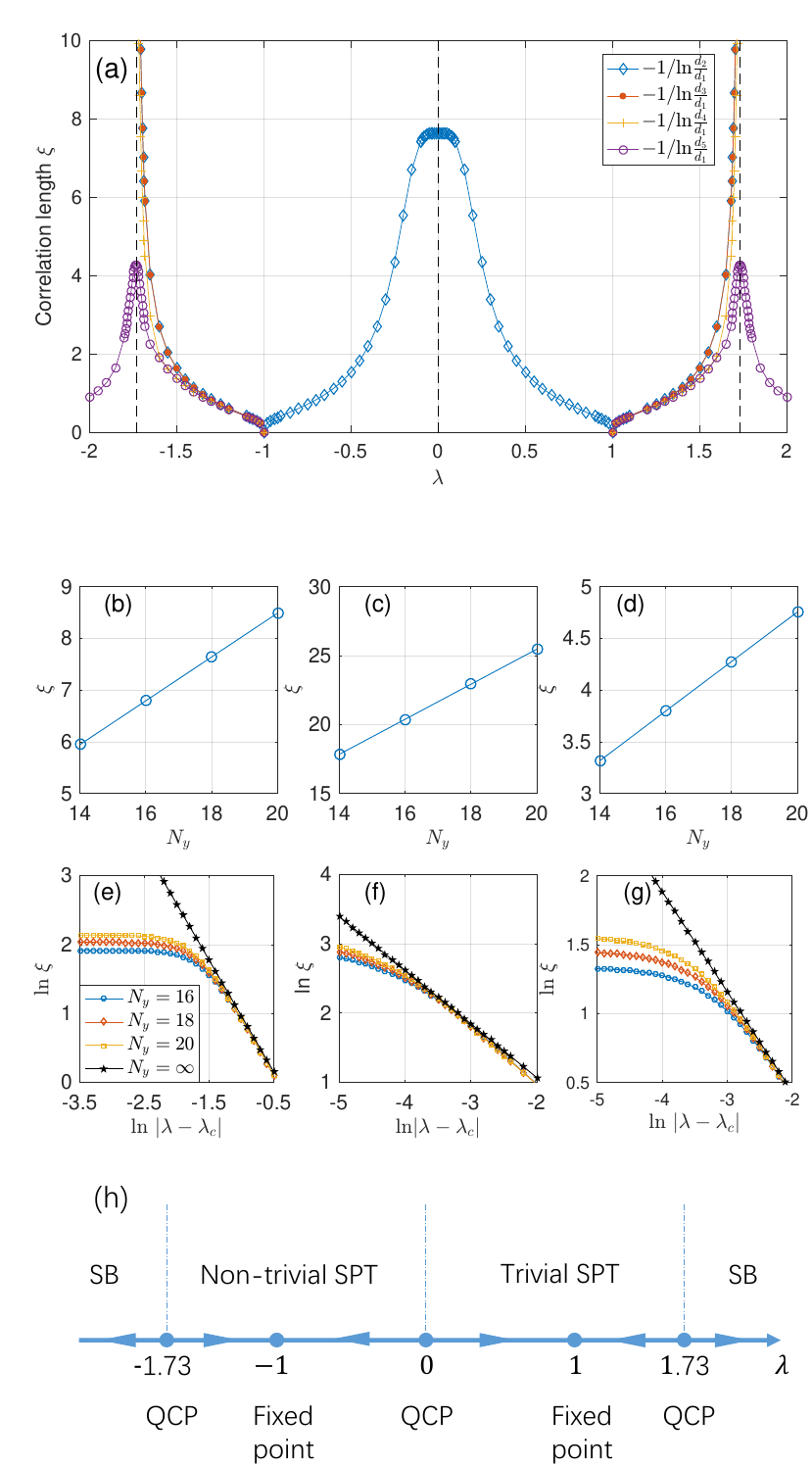}
\caption{(a) The quantities $-1/\text{ln}\frac{d_{i}}{d_{1}}$ of the SPT
wave function as function of $\protect\lambda $, where $d_{i}$ is the $i$-th
dominant eigenvalue of $\mathbb{T}_{0}$ and the circumference of the
transfer operator is $N_y=18$. (b), (c), (d) The correlation length $\protect%
\xi \propto N_{y}$ at $\protect\lambda =0$, around $|\protect\lambda %
|=1.73^{-}$, and $|\protect\lambda |=1.73^+$, respectively. (e),(f),(g) The
critical exponent $\protect\mu$ of $\protect\xi $ at $\protect\lambda%
\rightarrow 0$, $\protect\lambda \rightarrow 1.73^{-}$, and $\protect\lambda%
\rightarrow 1.73^{+}$, respectively. (h) The phase diagram is plotted and
several special points are marked, where the arrows denote the decreasing
direction of the correlation length. }
\label{correlation length}
\end{figure}

In the range that the dominant eigenvalue is four-fold degenerate, the
finite-size correlation length is defined by $\xi =-1/\text{ln}(|\frac{d_{5}%
}{d_{1}}|)$, where $d_i$ is the $i$-th dominant eigenvalues, while it is $%
\xi =-1/$ln$(|\frac{d_{2}}{d_{1}}|)$ in the range that the dominant
eigenvalue is non-degenerate. The calculated correlation length $\xi $ is
displayed in Fig.\ref{correlation length} for $|\lambda |\in \lbrack 0,2]$.
There exists three special points $\lambda_c=0$ and $\lambda_c =\pm 1.73$.
At $\lambda_c=0 $, the finite-size scaling of correlation length in Fig.\ref%
{correlation length}(b) leads to $\xi \varpropto N_{y}$, where $N_{y}$ is
the circumference of the transfer operator $\mathbb{T}_0$. So the
correlation length is divergent in the thermodynamic limit, corresponding to
a QCP. By carefully fitting the critical exponent of the correlation length in
the vicinity critical point with $\xi \sim |\lambda |^{-\nu }$, we can
extract the critical exponent $\nu \simeq 1.62$ shown in Fig.\ref%
{correlation length}(e). Since the correlation length is symmetric about $%
\lambda =0$, the critical exponent for both sides is the same.

However, close to $|\lambda _{c}|\simeq 1.73$, the finite-size correlation
length is defined by $\xi =-1/\text{ln}(\frac{d_{2}}{d_{1}})$ for $|\lambda
|<1.73$, while it is $\xi =-1/$ln$(\frac{d_{5}}{d_{1}})$ for $|\lambda
|>1.73 $. Fig.\ref{correlation length} (c) and (d) indicate that the
corresponding numerical values are proportional to $N_{y}$. Therefore, $%
\lambda \approx \pm 1.73$ correspond to two different QCPs and the critical
exponents of the correlation length for $|\lambda |\rightarrow 1.73^{-}$ and
$|\lambda |\rightarrow 1.73^{+}$ can also be determined as $\nu =0.77$ and $%
\nu ^{\prime }=0.73$, as shown in Fig.\ref{correlation length} (f) and (g).
At $\lambda =\pm 1$, the correlation length $\xi =0$ corresponds to the
stabilized fixed points of the SPT phases.

\section{Modular matrices of topologically ordered phases}

\label{Modular_Matrices}

The wave functions of topological states are given in Eq. (\ref%
{TOwavefunction}). Although the fixed point wave function of the double
semion model is different from the Levin-Gu wave function \cite{LevinGu}, it
can be demonstrated that, using the MPOs and modular matrices, it is indeed
the double semion state.

For topological states, the ground states have four-fold degeneracy on
torus, corresponding to the tensor network states without MPO inserting,
with MPO inserting in the horizontal direction, with MPO inserting in the
vertical direction, and with MPO inserting in both directions. The resulted
wave functions are $|\Psi (\mathbbm{1},\mathbbm{1})\rangle $, $|\Psi (%
\mathbbm{1},U_{\phi })\rangle $, $|\Psi (U_{\phi },\mathbbm{1})\rangle $ and
$|\Psi (U_{\phi },U_{\phi })\rangle $, where the winding numbers of domain
walls of $|\Psi (\mathbbm{1},\mathbbm{1})\rangle $ in the two directions are
both even, and inserting MPO changes the winding number parities of the
corresponding directions from even to odd. Notice that these wave functions
are not the minimally entangled states \cite{MES}, but are their linear
combinations.

\begin{figure}[tbp]
\includegraphics[width=9cm, trim=50 50 100 50, clip]{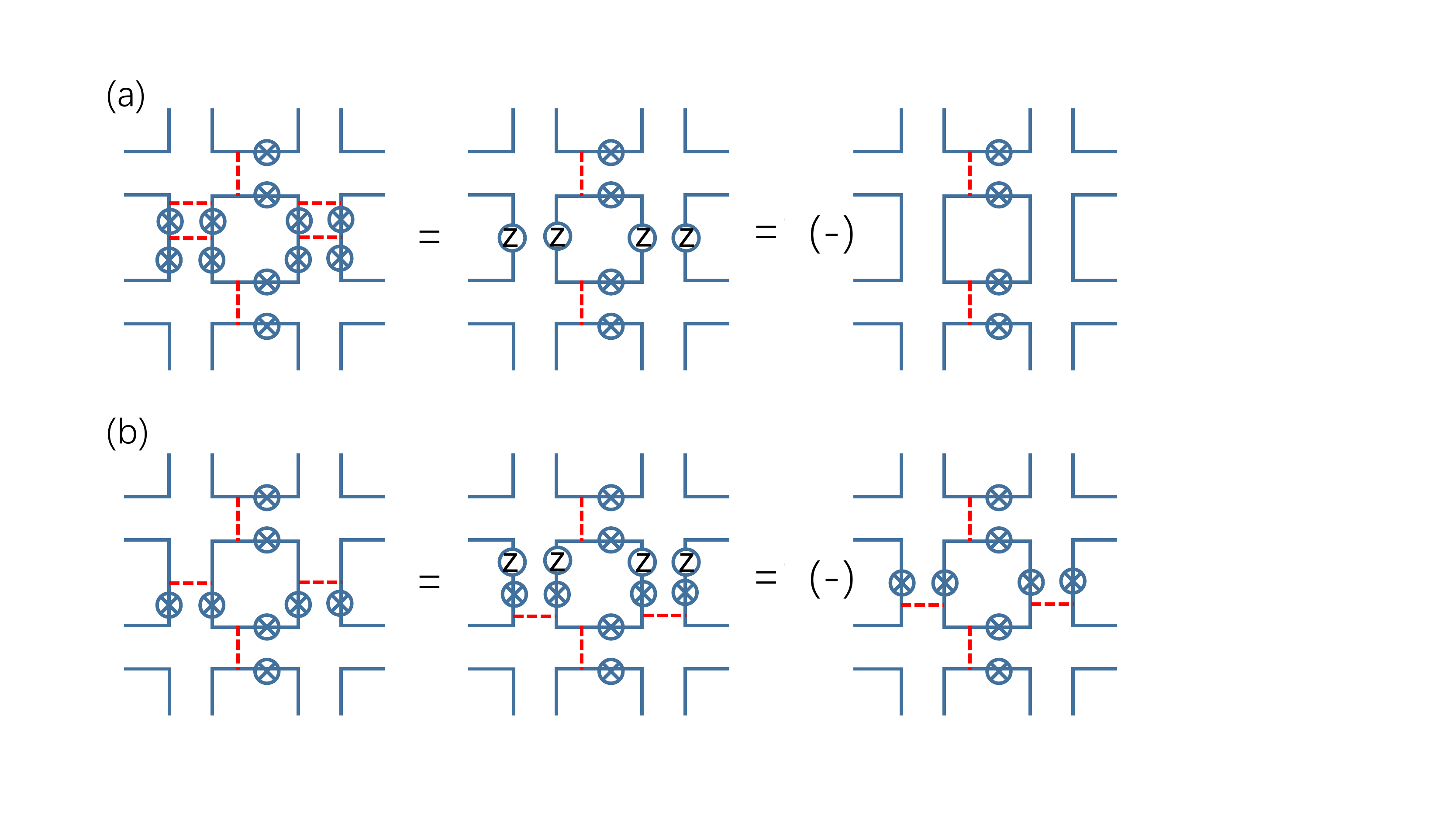}
\caption{(a) Composing two MPOs in the $x$ direction in the presence of a
MPO in the $y$ direction induces a minus sign in the $S$ matrix, where $X$
and $Z$ denote the Pauli matrices and the red dash lines are $CZ$ gates. (b)
Inversing the MPO in the $x$ direction in the presence of the MPO in the $y$
direction also gives rise to a minus sign in the $T$ matrix. }
\label{minus sign}
\end{figure}

In general, the only differences between the toric code and double semion
phases are the braiding statistics of anyons, which is encoded in the
modular matrices. There are two modular matrices: $S$ encodes the mutual
braiding statistics of anyons and $T$ encodes the self-braiding statistics.
Under the basis consisting of the wave functions $|\Psi (\mathbbm{1},%
\mathbbm{1})\rangle $, $|\Psi (\mathbbm{1},U_{\phi })\rangle $, $|\Psi
(U_{\phi },\mathbbm{1})\rangle $ and $|\Psi (U_{\phi },U_{\phi })\rangle $,
the modular matrices of the toric code model can be calculated as \cite%
{HuangWei1}
\begin{equation}
S=\left(
\begin{array}{cccc}
1 & 0 & 0 & 0 \\
0 & 0 & 1 & 0 \\
0 & 1 & 0 & 0 \\
0 & 0 & 0 & 1%
\end{array}%
\right) ,T=\left(
\begin{array}{cccc}
1 & 0 & 0 & 0 \\
0 & 1 & 0 & 0 \\
0 & 0 & 0 & 1 \\
0 & 0 & 1 & 0%
\end{array}%
\right) ,
\end{equation}%
and those for the double semion model are
\begin{equation}
S=\left(
\begin{array}{cccc}
1 & 0 & 0 & 0 \\
0 & 0 & 1 & 0 \\
0 & 1 & 0 & 0 \\
0 & 0 & 0 & -1%
\end{array}%
\right) ,T=\left(
\begin{array}{cccc}
1 & 0 & 0 & 0 \\
0 & 1 & 0 & 0 \\
0 & 0 & 0 & -1 \\
0 & 0 & 1 & 0%
\end{array}%
\right) .
\end{equation}%
The sign difference in $S$ and $T$ of the toric code model and double semion
model can be derived from the properties of MPOs. When there exists a MPO in
one direction for the double semion model, the combination of two MPOs or
the inverse the MPO in the other direction will give rise to a minus sign
\cite{GaugingTimeReversalSymmetry}, as shown in Fig.\ref{minus sign}. This
is resulted by the fact that the operators $CZ_{j,j+1}$ and $X_{j}X_{j+1}$
do not commute, i.e., $%
CZ_{j,j+1}X_{j}X_{j+1}=-X_{j}X_{j+1}CZ_{j,j+1}Z_{j}Z_{j+1}$. However, one
can easily verify that $\prod_{i}X_{i}^{\otimes 2}$ and $\prod_{i}CZ_{i,i+1}$
commute when acting on the periodic plaquettes and anti-commute when acting
on the twisted plaquettes (inserting a symmetry twist $X_{i}X_{i+1}$ in
another direction), as shown in Fig.\ref{minus sign}. The minus sign in $%
-Z_{i}Z_{i+1}$ is cancelled because we the considered systems have even
number of sites in both directions.

\section{Momenta related to topological spins}

\label{Momentum_polorization} A possible way to determine the momenta
related to of the eignstates is the momentum polarization \cite%
{MomentumPolarization}. But the momentum polarization for a single site
translation is not well-defined for the non-chiral topological states, so we
can only consider the well-defined translation of $2N_{y}$ sites ($4\pi $
rotation) \cite{SymmetryDefect}, and calculate the momentum polarization for
the translation of $2N_{y}$ sites and properly determine the anyon sectors.
As we mentioned in the Sec.III, the double layer transfer operator is $%
\mathbb{T}=\mathbb{T}_{0}\oplus \mathbb{T}_{1}$, where $\mathbb{T}$ has the $%
\mathbb{Z}_{2}\times \mathbb{Z}_{2}$ symmetry. Assuming that the dominant
eigenstate of $\mathbb{T}_{0}$ is $\sigma $ and that of $\mathbb{T}_{1}$ is $%
\sigma ^{\prime }$ and considering the $U_{\text{czx}}\otimes \mathbbm{1}$
symmetry is broken, we then have
\begin{equation}
\sigma =U_{\text{czx}}\sigma ^{\prime }\mathbbm{1},\sigma ^{\prime }=U_{%
\text{czx}}\sigma \mathbbm{1}.
\end{equation}%
The similar arguments hold for $\mathbb{T}_{\text{czx}}^{\text{czx}}$ and
\begin{equation}
\sigma _{\text{czx}}\ =-U_{\text{czx}}\sigma _{\text{czx}}^{\prime }%
\mathbbm{1},\sigma _{\text{czx}}^{\prime }=U_{\text{czx}}\sigma _{\text{czx}}%
\mathbbm{1},
\end{equation}%
where $\sigma _{\text{czx}}$ and $\sigma _{\text{czx}}^{\prime }$ are the
dominant eigenstates of $\mathbb{T}_{\text{0, czx}}^{\text{czx}}$ and $%
\mathbb{T}_{\text{1, czx}}^{\text{czx}}$. Although the dominant eigenstates
of $\mathbb{T}$ and $\mathbb{T}_{\text{czx}}^{\text{czx}}$ break the
symmetries $\mathbbm{1}\otimes U_{\text{czx}}\ $ and $U_{\text{czx}}\otimes %
\mathbbm{1}$, we can restore these symmetries by linearly combining the
dominant eigenstates \cite{PEPSTO}, because the reduced density matrix can
be linear combinations of $\sigma $, $\sigma ^{\prime }$, $\sigma _{\text{czx%
}}$ and $\sigma _{\text{czx}}^{\prime }$. So we have the symmetric dominant
eigenstates
\begin{eqnarray}
\rho &=&\sigma +\sigma ^{\prime },\ \rho ^{\prime }=\sigma -\sigma ^{\prime }
\notag \\
\rho _{\text{czx}} &=&\sigma _{\text{czx}}+i\sigma _{\text{czx}}^{\prime },\
\rho _{\text{czx}}^{\prime }=\sigma _{\text{czx}}-i\sigma _{\text{czx}%
}^{\prime },
\end{eqnarray}%
which satisfies
\begin{eqnarray}
U_{\text{czx}}\rho \mathbbm{1} &=&\rho ,\ U_{\text{czx}}\rho ^{\prime }%
\mathbbm{1}=-\rho ^{\prime },  \notag \\
U_{\text{czx}}\rho _{\text{czx}}\mathbbm{1} &=&-i\rho _{\text{czx}},\ U_{%
\text{czx}}\rho _{\text{czx}}^{\prime }\mathbbm{1}=i\rho _{\text{czx}%
}^{\prime }.
\end{eqnarray}%
By considering the constraints
\begin{equation}
T^{2N_{y}}=\mathbbm{1}, \tilde{T}^{2N_{y}}=U_{\text{czx}}^{2},
\end{equation}
the momentum polarizations with $2N_{y}$ site translation are thus given by
\begin{eqnarray}
\text{tr}(\mathbbm{1}\rho ) &=&1,\ \text{tr}(\mathbbm{1}\rho ^{\prime })=1,
\notag \\
\text{tr}(U_{\text{czx}}^{2}\rho _{\text{czx}}) &=&-1,\ \text{tr}(U_{\text{%
czx}}^{2}\rho _{\text{czx}}^{\prime })=-1,
\end{eqnarray}%
which means that the $4\pi $ rotation acquires a minus sign for the sectors
with flux and the topological spins are $\pm i$. Therefore, the momenta of
excitations in the semion and anti-semion sectors should have a correction.

\bibliography{ref}

\end{document}